\documentclass[aps,preprintnumbers,preprint,nofootinbib,12pt]{revtex4}
\usepackage{graphicx,amsfonts,amssymb}
\usepackage[mathscr]{euscript}
\usepackage{graphics,epsfig, psfrag}
\usepackage{graphics,epsfig, psfrag}
\usepackage{amsmath}
\usepackage{amscd}
\usepackage{amssymb}
\usepackage{amsfonts}
\usepackage{eufrak}
\usepackage{fancyheadings}
\usepackage[mathscr]{euscript}
\usepackage{pstricks}
\usepackage{pst-node}

\makeatletter       
\renewcommand{\thesection}{\Roman{section}}
\renewcommand{\p@subsection}{}

\renewcommand{\p@subsubsection}{}

\makeatother
\makeatletter
\@addtoreset{equation}{section}
\makeatother
\renewcommand{\theequation}{\arabic{section}.\arabic{equation}}
\begin{document}

\pagestyle{plain}

\title{\vspace{.5cm}
\Large Hilbert Space Structure in Classical Mechanics: (II)}

\author{\large E. Deotto}
\email[email: ]{deotto@mitlns.mit.edu}
\affiliation{Center for Theoretical Physics,
             Massachusetts Institute of Technology, \\
	     Cambridge, MA 02139, USA}

\author{\large E. Gozzi}
\email[email: ]{gozzi@ts.infn.it}

\author{\large D. Mauro}
\email[email: ]{mauro@ts.infn.it}
\affiliation{Dipartimento di Fisica Teorica, \\
	     Universit\`a di Trieste, 
	     Strada Costiera 11, P.O.Box 586, Trieste, Italy,
	     and INFN, Sezione di Trieste, Italy}

\date[]{7 August 2002}
\preprint{MIT-CTP-3294}

\begin{abstract}
In this paper we analyze two different functional formulations of classical
mechanics. In the first one the Jacobi fields are represented by {\it bosonic}
variables and belong to the vector (or its dual) representation of the symplectic 
group. In the second formulation the Jacobi fields are given as {\it condensates}
of Grassmannian variables belonging to the spinor representation of the metaplectic 
group. For both formulations we shall show that, differently from what happens
in the case presented in paper no. (I), it is possible to endow the associated Hilbert space
with a positive definite scalar product and to describe the dynamics via a Hermitian Hamiltonian.
The drawback of this formulation is that higher forms do not appear automatically 
and that the description of chaotic systems may need a further extension of the Hilbert 
space.

\end{abstract}
\maketitle
--------------------------------------------------------------------------------
\section{Introduction}

In a previous paper with the same title \cite{one} we analyzed in detail
the Hilbert space structure associated to the {\it standard} path integral
formulation \cite{Gozzi} of classical mechanics ({\it CM}). We called {\it standard}
formulation the one in which the Jacobi fields \cite{Gozzi} are represented by
{\it Grassmannian} variables and belong to the vector (or its dual)
representation of the symplectic group. 
In Ref. \cite{one} we showed that the associated Hilbert
space cannot have at the same time a positive definite scalar product and a
Hermitian Hamiltonian. We shall indicate this formulation as {\it CPI} for
``{\underline C}lassical {\underline P}ath {\underline I}ntegral".

In {\bf Sec. II} of this paper we will review a different functional approach
\cite{regini} to {\it CM} in which the Jacobi fields are represented by {\it bosonic}
variables, instead of {\it Grassmannian} ones, but they still belong to the 
vector (or its dual) representation of the symplectic group. We will
indicate this formulation as {\it BFA} for {\underline B}osonic {\underline F}unctional
{\underline A}pproach. The operatorial version of
the {\it BFA} is studied in detail in {\bf Sec. III}. There
we will prove that it is possible to have both a
positive definite Hilbert space and a Hermitian Hamiltonian differently from
what happens in the {\it CPI} case \cite{one}. In {\bf Sec. IV} we shall present a
geometrical analysis of the {\it BFA} interpreting the various
variables as basis for the forms and vector fields. Like in the {\it CPI}
several symmetries make their appearance. The analysis of these symmetries
requires, in the bosonic case, a special care. 
A special care requires also the construction of higher forms whose study is
performed in {\bf Sec. V}. Their construction is less straightforward than in
the {\it CPI} case but it has its own consistency. 

Both in the {\it CPI}
and in the {\it BFA} case, the Jacobi fields belong to the vector (or its dual)
representation of the symplectic group, but this is not the only 
representation we can use. In
fact in {\bf Sec. VI} we will review another functional approach to {\it CM}
in which the Jacobi fields are built as condensates of Grassmannian
variables belonging to the spinor representation of the {\it metaplectic}
group \cite{meta}. We shall indicate this formulation as {\it MFA} for {\underline M}etaplectic
{\underline F}unctional {\underline A}pproach. 
We shall show in {\bf Sec. VII} that
also in the {\it MFA} case, like in the {\it BFA} one, it is possible to construct both a
positive definite Hilbert space and a Hermitian Hamiltonian. 

In the {\bf Conclusions},
besides drawing an overall picture from the technical analysis contained in 
this paper and of Ref. \cite{one},
we explain why the problems (non-hermiticity, etc.) apparently by-passed by the {\it BFA}
and {\it MFA} with respect to the {\it CPI} actually lead to other problems whose solution 
could lie in a further extension of the Hilbert space \cite{benatti}. 

Many calculations of Ref. \cite{regini} and \cite{meta} are reproduced here in
detail to make this paper self-contained. Some of these and other details are
confined to few Appendices. 

\section{Bosonic Path Integral}

The {\it CPI} formulation of {\it CM} \cite{Gozzi} has, as starting point, the
following generating functional 
\begin{equation}
\displaystyle Z[J]=\int {\mathscr D}\varphi \,
\widetilde{\delta}[\varphi^a(t)-\varphi^a_{cl}(t)] \,\text{exp} \biggl[ \int dt \,J_a\varphi^a\biggr].
\label{2-1}
\end{equation}
The variables $\varphi^a$ are the phase space coordinates: $\varphi^a\equiv
(q^i,p^i)$ of a symplectic manifold ${\cal M}$ with $a=1,\ldots,2n$ and
$i=1,\ldots,n$, and $\varphi^a_{cl}(t)$ are the solutions of the Hamiltonian
equations of motion
\begin{equation}
\displaystyle \dot{\varphi}^a=\omega^{ab}\frac{\partial H}{\partial \varphi^b}
\end{equation}
with $\omega^{ab}$ the standard symplectic matrix. Disregarding for a moment the
current $J_a$, it is easy to realize that we can write $Z[J]$ in (\ref{2-1})
as
\begin{equation}
\displaystyle Z[J]=\int {\cal
D}\varphi\,\widetilde{\delta}\biggl[\dot{\varphi}^a-\omega^{ab}\frac{\partial
H}{\partial
\varphi^b}\biggr]\text{det}\biggl[\delta_l^a\partial_t-\omega^{ab}\frac{\partial^2H}{\partial
\varphi^b\partial\varphi^l}\biggr] \label{2-2}
\end{equation}
where the determinant appearing in (\ref{2-2}) is a functional determinant 
which is needed
to pass from the zeroes (in (\ref{2-1})) of the function 
$\displaystyle F(\varphi,\dot{\varphi})\equiv \dot{\varphi}^a-\omega^{ab}\frac{\partial H}
{\partial \varphi^b}$ to the function itself in (\ref{2-2}).
This functional determinant is positive definite \cite{schulman} and this {\it crucial}
property is based on the fact that between two phase space points there is at most only
one trajectory. This property does not hold between two points of configuration space
and so the associated functional determinant is not positive definite.

In the {\it CPI} formulation \cite{Gozzi} of {\it CM} the next step was to 
``exponentiate" the determinant in (\ref{2-2}) via Grassmannian variables like it is done 
in the Faddeev-Popov ({\it FP}) method of gauge theories. In Ref. \cite{regini} we chose a
different strategy. The trick we adopted was to substitute the determinant in (\ref{2-2})
with an inverse determinant:
\begin{equation}
\displaystyle \text{det}\biggl[\delta_l^a\partial_t-\omega^{ab}\frac{\partial^2H}{\partial
\varphi^b\partial\varphi^l}\biggr]=\biggl\{\text{det}\biggl[\delta_l^a\partial_t+\omega^{ab}
\frac{\partial^2H}{\partial\varphi^b\partial\varphi^l}\biggr]\biggr\}^{-1}
\label{2-3}
\end{equation}
and the proof of this relation is given in Appendix A. The next step done in Ref. \cite{regini}
was to use (\ref{2-3}) in 
(\ref{2-2}) and then ``exponentiate" the inverse matrix via bosonic variables using 
the well-known formula
\begin{equation}
\int dx^idy_j \, \text{exp}\,(ix^iA^j_iy_j) \propto \{\text{det}[A_i^j]\}^{-1}. \label{2-4}
\end{equation}
This formula of Gaussian integration applies only to matrices with positive determinant and this is 
our case as we explained above. Note that this is no longer the case for the {\it FP} 
determinant which, as signalled by the Gribov problem, is not positive definite. This is the
reason why the {\it FP} determinant could not be exponentiated via bosonic variables. Various attempts
exist in the literature to write fermionic determinants via bosonic variables \cite{slavnov} but they
are all different from the one we have presented here.

Let us now use the relations (\ref{2-3}) and (\ref{2-4}) into the $Z[J]$ of (\ref{2-2}), the result is
\begin{eqnarray}
\displaystyle Z_{\scriptscriptstyle BFA}[J]&=&\int {\mathscr D} \varphi^a\widetilde{\delta}\biggl[\dot{\varphi}^a-\omega^{ab}
\frac{\partial H}{\partial \varphi^b}\biggr]\biggl\{\text{det}\biggl[\delta^a_l\partial_t
+\omega^{ab}\frac{\partial^2H}{\partial\varphi^b\partial\varphi^l}\biggr]\biggr\}^{-1} \nonumber\\
&=&\int {\mathscr D}\varphi^a{\mathscr D}\lambda_a{\mathscr D}\pi^a{\mathscr D}\xi_a\,\text{exp}\,
\biggl(i\int dt \,{\cal L}_{\scriptscriptstyle BFA}\biggr)
\label{2-5-a}
\end{eqnarray}
where 
\begin{equation}
{\cal L}_{\scriptscriptstyle BFA}=\lambda_a\biggl[\dot{\varphi}^a-\omega^{ab}\frac{\partial H}{\partial \varphi^b}\biggr]+\pi^l
\biggl[\delta_l^a\partial_t+\omega^{ab}\frac{\partial^2H}{\partial \varphi^b\partial\varphi^l}\biggr]\xi_a.
\label{2-5-b}
\end{equation}
The variables $\lambda_a$ are the same as in the {\it CPI} \cite{Gozzi} formulation of {\it CM}
and are needed to produce a Fourier transform of the Dirac delta $\displaystyle \widetilde{\delta}\biggl(\dot{\varphi}^a-
\omega^{ab}\frac{\partial H}{\partial\varphi^b}\biggr)$. The variables $\lambda_a$ are {\it bosonic}
like $\pi^l,\xi_a$ which were introduced to exponentiate the matrix 
$\displaystyle \biggl[\delta_l^a\partial_t+\omega^{ab}\frac{\partial^2H}{\partial\varphi^b\partial\varphi^l}
\biggr]$ and produce the inverse determinant of (\ref{2-5-a}). The $\pi,\xi$ are the analogs of the 
$x^i,y_j$ variables of (\ref{2-4}). In the {\it CPI} formulation of {\it CM} 
\cite{Gozzi} the Lagrangian obtained was
\begin{equation}
\widetilde{\cal L}_{\scriptscriptstyle CPI}=\lambda_a\biggl[\dot{\varphi}^a-\omega^{ab} \frac{\partial H}{\partial\varphi^b}
\biggr]+i\bar{c}_a\biggl[\delta_l^a\partial_t
-\omega^{ab}\frac{\partial^2H}{\partial\varphi^b\partial\varphi^l}\biggr]c^l \label{2-6}
\end{equation}
which can be compared with ${\cal L}_{\scriptscriptstyle BFA}$ of (\ref{2-5-a}) if, in this last one, we interchange
$\pi^l,\xi_a$. The result is
\begin{equation}
{\cal L}_{\scriptscriptstyle BFA}=\lambda_a\biggl[\dot{\varphi}^a-\omega^{ab}\frac{\partial H}{\partial \varphi^b}\biggr]-
\xi_a\biggl[\delta_l^a\partial_t-
\omega^{ab}\frac{\partial^2H}{\partial\varphi^b\partial\varphi^l}\biggr]\pi^l+(\text{s.t.})
\label{2-7}
\end{equation}
where (s.t.) is a surface term.
From (\ref{2-7}) we see that, modulo the surface term, we get the ${\cal L}_{\scriptscriptstyle BFA}$
from $\widetilde{\cal L}_{\scriptscriptstyle CPI}$ by replacing 
the Grassmannian variables $i\bar{c}_a$ and $c^l$ with
the bosonic ones $-\xi_a$ and $\pi^l$.

\section{Operatorial Formalism}

We should now build the operatorial formalism associated to the {\it BFA}. 
The commutators among the basic variables ($\varphi^a,\lambda_a,\pi^a,\xi_a$) can be straightforwardly
derived from the path integral (\ref{2-5-a}) by inspecting the kinetic term in (\ref{2-5-b}). They turn
out to be 
\begin{eqnarray}
&&[\widehat{\varphi}^a,\widehat{\lambda}_b]=i\delta_b^a \nonumber\\
&&[\widehat{\xi}_a,\widehat{\pi}^b]=i\delta_a^b \label{3-1}
\end{eqnarray}
where we have now turned the path integral variables into operators. Next we choose the ``Schr\"odinger" 
representation in which $\widehat{\varphi}^a$ and $\widehat{\pi}^a$ are realized as multiplicative operators while 
$\widehat{\lambda}_a$ and $\widehat{\xi}_a$ as derivative ones of the 
form: 
\begin{equation}
\left\{
	\begin{array}{l}
	\displaystyle \label{3-2}
	\widehat{\lambda}_a\equiv-i\frac{\partial}{\partial\varphi^a} \smallskip \\
	\displaystyle \widehat{\xi}_a\equiv i\frac{\partial}{\partial\pi^a}.
	\end{array}
	\right.
\end{equation}
So in this representation the associated Hilbert space is made of ``wave functions" $\psi(\varphi^a,\pi^a)$ on the
$4n$-dimensional ``configurational" space whose coordinates are $(\varphi^a,\pi^a)$. A very natural, and 
{\it positive definite}, scalar product that we can introduce in this space is
\begin{equation}
\displaystyle \langle \psi|\psi^{\prime}\rangle\equiv \int d^{2n}\varphi^ad^{2n}\pi^a\,\psi^*(\varphi^a,\pi^a)
\psi^{\prime}(\varphi^a,\pi^a). \label{3-3}
\end{equation}
It is extremely easy to check that the $8n$ operators $\widehat{\varphi}^a,\widehat{\lambda}_a,
\widehat{\pi}^a,\widehat{\xi}_a$ are all Hermitian under the scalar product (\ref{3-3})
\begin{equation}
\left\{
	\begin{array}{l}
	\displaystyle \label{3-4}
	\widehat{\varphi}^{a\dagger}=\widehat{\varphi}^a \smallskip \\
	\widehat{\lambda}_a^{\dagger}=\widehat{\lambda}_a \smallskip \\
	\widehat{\xi}_a^{\,\dagger}=\widehat{\xi}_a \smallskip \\
	\widehat{\pi}^{a\dagger}=\widehat{\pi}^a.
	\end{array}
	\right.
\end{equation}
Let us now turn to the Hamiltonian which can be easily derived from the Lagrangian in (\ref{2-5-b}) 
\begin{equation}
{\cal H}_{\scriptscriptstyle BFA}=\lambda_a\omega^{ab}\partial_bH-\pi^l\omega^{ab}\partial_b\partial_lH\xi_a. \label{3-5-a}
\end{equation}
Turning the variables ($\varphi^a,\lambda_a,\pi^a,\xi_a$) into operators, the Hamiltonian itself becomes
\begin{equation}
\widehat{\cal H}_{\scriptscriptstyle BFA}=\widehat{\lambda}_a\omega^{ab}\partial_bH-\widehat{\pi}^l\omega^{ab}\partial_b\partial_lH\widehat{\xi}_a.
\label{3-5}
\end{equation}
It is straightforward to check that this Hamiltonian is Hermitian under the hermiticity conditions (\ref{3-4})
\begin{eqnarray}
\widehat{\cal H}_{\scriptscriptstyle BFA}^{\dagger}&=&(\widehat{\lambda}_a\omega^{ab}\partial_bH-\widehat{\pi}^l\omega^{ab}\partial_b\partial_lH
\widehat{\xi}_a)^{\dagger}=\nonumber\\
&=&(\partial_bH)^{\dagger}\omega^{ab}\widehat{\lambda}_a^{\dagger}
-\widehat{\xi}_a^{\dagger}\omega^{ab}(\partial_b\partial_lH)^{\dagger}\widehat{\pi}^{l\dagger}=\nonumber\\
&=&(\partial_bH)\omega^{ab}\widehat{\lambda}_a-\widehat{\xi}_a\omega^{ab}\partial_b\partial_lH\widehat{\pi}^l=\nonumber\\
&=&\widehat{\lambda}_a\omega^{ab}\partial_bH+i\omega^{ab}\partial_a\partial_bH
-\widehat{\pi}^l\omega^{ab}\partial_b\partial_lH\widehat{\xi}_a-i\delta_a^l
\omega^{ab}\partial_b\partial_lH=\nonumber\\
&=&\widehat{\cal H}_{\scriptscriptstyle BFA}. \label{3-6}
\end{eqnarray}
In the last two steps, in order to exchange $\widehat{\lambda}_a$ with $\partial_bH$ and $\widehat{\pi}^l$ with 
$\widehat{\xi}_a$, 
we have used their commutation relations $[\widehat{\lambda}_a,\partial_bH]=-i\partial_a\partial_bH$, 
$[\widehat{\xi}_a,\widehat{\pi}^l]=i\delta_a^l$ and the fact that $\omega^{ab}\partial_b\partial_aH$ is zero because of
the antisymmetry of $\omega^{ab}$. So we can conclude by saying that, differently than in the {\it CPI} 
case analyzed in Ref. \cite{one}, in the
{\it BFA} case we can have both a positive definite Hilbert space and a Hermitian Hamiltonian. 

The reader may remember that in Ref. \cite{one} we gave some {\it physical} reasons of why we could not have both a positive 
definite Hilbert space and a Hermitian Hamiltonian in the {\it CPI} case. So it is crucial to find out how we can 
bypass those {\it physical} reasons in the {\it BFA}. Those reasons (explained in the conclusions of Ref. \cite{one})
were basically the following: in a chaotic system the Jacobi fields $c^a(t)$ grow exponentially and as a consequence some 
of the wave functions of the form 
\begin{equation}
\psi=\psi_ac^a \label{3-7}
\end{equation}
have a norm which also grows exponentially with time (see Appendix E of Ref. \cite{one}). This means that the norm is not
conserved and for this to happen we need a non-unitary evolution or equivalently a non Hermitian Hamiltonian.
This kind of reasoning cannot be applied in the {\it BFA} case. In fact here the role of the Jacobi fields is taken by the 
variables $\pi^a$ whose equations of motion can be derived from the Lagrangian ${\cal L}_{\scriptscriptstyle BFA}$ 
of (\ref{2-7}):
\begin{equation}
\dot{\pi}^a=\omega^{ad}\partial_d\partial_bH\pi^b, \label{3-8-b}
\end{equation}
and so the analog of the wave function (\ref{3-7}) is:
\begin{equation}
\psi(\varphi,\pi)=\psi_a(\varphi)\pi^a. \label{3-8}
\end{equation}
Unfortunately this wave function is not normalizable according to the scalar product (\ref{3-3}). So even if the exponential 
increase in $\pi^a$ would imply, like for the wave function (\ref{3-7}), an exponential increase of the norm of the state, 
this would not lead to the conclusion that the evolution is not unitary. 
The reason is that the state (\ref{3-8}) 
itself is not part of the Hilbert space already at $t=0$ (it is not normalizable)
and consequently the Hamiltonian would not act on it. If
the reader is not immediately  convinced by our arguments, he should remember that the line of reasoning we 
followed 
in the conclusions and in Appendix E of Ref. \cite{one} to motivate our physical understanding, was crucially based on the use 
of wave functions linear in the Jacobi fields. These no longer belong to our new Hilbert space.

\section{Geometrical Analysis and Symmetries}

In Ref. \cite{regini} we tried a first geometrical analysis of the bosonic formalism. There we gave an interpretation
of the variables $\pi^a,\xi_a$ as {\it components} of vectors and forms whose {\it basis} were respectively
the variables $\bar{c}_a$ and $c^a$:
\begin{equation}
\left\{
	\begin{array}{l}
	\displaystyle \label{4-1}
	V=\pi^a\bar{c}_a \smallskip\\
	F=\xi_ac^a.
	\end{array}
	\right.
\end{equation}
The reason was that, under an infinitesimal diffeomorphism generated by ${\cal H}_{\scriptscriptstyle BFA}$ 
over the original phase space:
\begin{equation}
\varphi^{\prime a}=\varphi^a+\epsilon\omega^{ab}\partial_bH,
\end{equation}
the variables $\pi^a$ and $\xi_a$ transform in the following way:
\begin{eqnarray}
\displaystyle &&\pi^{a\prime}=\pi^a+\epsilon\omega^{ac}\partial_c\partial_bH\pi^b=
\frac{\partial \varphi^{a\prime}}{\partial\varphi^b}\pi^b\nonumber\\
&&\xi_a^{\prime}=\xi_a-\epsilon\omega^{bc}\partial_c\partial_aH\xi_b=
\frac{\partial\varphi^b}{\partial\varphi^{a\prime}}\xi_b,
\end{eqnarray}
i.e. just as components of vectors and forms.
In this interpretation 
the Hamiltonian $\widehat{\cal H}_{\scriptscriptstyle BFA}$ of (\ref{3-5}) cannot be given the meaning of a Lie derivative.
In fact we know that a Lie derivative, ${\cal L}_{(dH)^{\sharp}}$ \cite{marsden}, changes the components of a vector 
as follows:
\begin{equation}
\delta \pi^l=(\partial_a\pi^l)\omega^{ab}\partial_bH-(\partial_a\omega^{lb}\partial_bH)\pi^a \label{4-2}
\end{equation}
while $\widehat{\cal H}_{\scriptscriptstyle BFA}$ 
in (\ref{3-5}) induces on $\widehat{\pi}^l$ the following transformation
\begin{equation}
\delta \widehat{\pi}^l=[\widehat{\pi}^l,i\widehat{\cal H}_{\scriptscriptstyle BFA}]=
(-\partial_a\omega^{lb}\partial_bH)\widehat{\pi}^a \label{4-3}
\end{equation}
which is clearly different from (\ref{4-2}).
So if we insist in the analysis presented in Ref. \cite{regini}, we should {\bf first} abandon the interpretation
of $\widehat{\cal H}$ of the {\it BFA} case as the Lie derivative along the Hamiltonian flow.
{\bf Second}, if we insist in interpreting $\pi^a,\xi_a$ as components, we should make them dependent on $\varphi$ 
and that implies that we should give a connection to glue the fibers of $T^*(T^*{\cal M})$ of which 
$\pi^a$ and $\xi_a$ are coordinates \cite{regini}. This connection does not appear in a natural way in our formalism.
So, in order to bypass these {\it two} problems, we will interpret $\widehat{\pi}^a$ and $\widehat{\xi}_a$ as 
{\it basis} respectively of forms and vector fields. 
One-forms and vector fields are then given by 
\begin{equation}
\left\{
	\begin{array}{l}
	\displaystyle \label{4-4}
	F=F_a(\widehat{\varphi})\widehat{\pi}^a \smallskip\\
	V=V^a(\widehat{\varphi})\widehat{\xi}_a.
	\end{array}
	\right.
\end{equation}
As a consequence, it is easy to check that the $\widehat{\cal H}_{\scriptscriptstyle BFA}$ 
of (\ref{3-5}) can be interpreted as 
the Lie derivative (up to a constant factor). 
In fact, applying $i\widehat{\cal H}_{\scriptscriptstyle BFA}$ to the vector $V$ of (\ref{4-4}), we get
\begin{equation}
[i\widehat{\cal H}_{\scriptscriptstyle BFA}, V^e(\widehat{\varphi})\widehat{\xi}_e]=[(\partial_aV^e)\omega^{ab}\partial_bH-(\partial_a\omega^{eb}
\partial_bH)V^a]\widehat{\xi}_e \label{4-5}
\end{equation}
and this is exactly how vector components $V^a$ change \cite{marsden} under the Lie derivative of the Hamiltonian flow:
\begin{equation}
\delta V^e(\varphi)\equiv V^{e\prime}(\varphi)-V^e(\varphi)=
(\partial_aV^e)\omega^{ab}\partial_bH-(\partial_a\omega^{eb}\partial_bH)V^a. \label{4-6}
\end{equation}
Analogously, on the one-forms $F=F_a(\widehat{\varphi})\widehat{\pi}^a$ of (\ref{4-4}), $\widehat{\cal
H}_{\scriptscriptstyle BFA}$ acts as follows:
\begin{equation}
[i\widehat{\cal H}_{\scriptscriptstyle BFA}, F_e(\widehat{\varphi})\widehat{\pi}^e]=[(\partial_aF_e)\omega^{ab}\partial_bH+(\partial_e\omega^{ab}\partial_bH)
F_a]\pi^e. \label{4-7}
\end{equation}
This is exactly how one-forms transform \cite{marsden} under the Lie derivative:
\begin{equation}
\delta F_e(\varphi)=F_e^{\prime}(\varphi)-F_e(\varphi)=(\partial_aF_e)\omega^{ab}\partial_bH+
(\partial_e\omega^{ab}\partial_bH)F_a. \label{4-8}
\end{equation}
To give to $\widehat{\cal H}_{\scriptscriptstyle BFA}$ the meaning of
a Lie derivative, another check we should do is the following. The
commutator of two Lie derivatives has the property \cite{marsden}:
\begin{equation}
\displaystyle [{\cal L}_{(dH_1)^{\sharp}},{\cal L}_{(dH_2)^{\sharp}}]={\cal L}_{[(dH_1)^{\sharp},(dH_2)^{\sharp}]_{Lb}}
\label{4-9}
\end{equation}
where $H_1$ and $H_2$ are the Hamiltonians entering the Lie derivative and $[(dH_1)^{\sharp},(dH_2)^{\sharp}]_{Lb}$ 
are the {\it Lie brackets} (Lb) between the associated Hamiltonian vector fields. According
to our conventions the Lie brackets can be related to the Poisson brackets between $H_{\scriptscriptstyle 1}$ and 
$H_{\scriptscriptstyle 2}$ in the following way \cite{marsden}:
\begin{eqnarray}
\displaystyle [(dH_{\scriptscriptstyle 1})^{\sharp},(dH_{\scriptscriptstyle 2})^{\sharp}]_{Lb}&=&
[\omega^{bc}\partial_cH_{\scriptscriptstyle 1}\partial_b\omega^{ad}\partial_dH_{\scriptscriptstyle 2}
-\omega^{bc}\partial_cH_{\scriptscriptstyle 2}(\partial_b\omega^{ad}\partial_dH_{\scriptscriptstyle 1})]\xi_a=
\nonumber\\
&=&-[\omega^{ad}\partial_d(\partial_bH_{\scriptscriptstyle 1}\omega^{bc}\partial_cH_{\scriptscriptstyle 2})]\xi_a=
-(d\{H_{\scriptscriptstyle 1},H_{\scriptscriptstyle 2}\})^{\sharp} \label{correspondence}.
\end{eqnarray}
Therefore (\ref{4-9}) can be rewritten as:
\begin{equation}
\displaystyle [{\cal L}_{(dH_1)^{\sharp}},{\cal L}_{(dH_2)^{\sharp}}]=
{\cal L}_{-(d\{H_{\scriptscriptstyle 1},H_{\scriptscriptstyle 2}\})^{\sharp}}. \label{4-9-bis}
\end{equation}
As we associate to each Lie derivative ${\cal L}_{(dH)^{\sharp}}$ an operator 
$i\widehat{\cal H}_{\scriptscriptstyle H}$,
the relation (\ref{4-9-bis}) should turn into the following one
\begin{equation}
[i\widehat{\cal H}_{H_1},i\widehat{\cal H}_{H_2}]=-i\widehat{\cal H}_{\{H_1,H_2\}} \label{4-10}
\end{equation}
where we have put on the $\widehat{\cal H}$ of (\ref{3-5}), the label $H_1$, $H_2$ or $\{H_1,H_2\}$ to indicate
the function entering each $\widehat{\cal H}_{\scriptscriptstyle BFA}$.
It is very easy to verify (\ref{4-10}) and this is done in detail in Appendix B. 

In the {\it CPI} \cite{Gozzi} it was found that there were various 
conserved universal charges associated to the Lie derivative. They were called \cite{Gozzi} BRS, anti-BRS, ghost and supersymmetry charges in analogy with similar 
objects present in gauge field theory. Despite these names they are well-known structures in symplectic geometry 
\cite{marsden}; for example the BRS charge is nothing but the exterior derivative on phase space and its conservation
is related to the fact that the exterior derivative commutes with the Lie derivative \cite{marsden}. The ghost charge 
is basically the form number \cite{marsden} while the supersymmetry charge is connected to the concept of equivariant 
cohomology \cite{deotto}. Besides their geometrically universal meaning, these charges and the associated symmetries
somehow signal the redundancy of the $8n$ variables $(\varphi^a,\lambda_a,c^a,\bar{c}_a)$ used in 
describing {\it CM}. We know in fact that {\it CM} can be described using only the $2n$ phase space variables ($\varphi^a$) and so the
other $6n$ variables must be related to the $\varphi^a$ via some symmetries which should be present for any system. Also
in the bosonic case we have many extra variables $(\lambda_a,\pi^a,\xi_a)$ besides the $2n$ phase space variables
$\varphi^a$ and so we expect to find various symmetries like in the {\it CPI}. 

The way we start our search for the symmetries in the bosonic case is rather naive but it is one of the few 
we could think of. Basically, as the variables $\pi^a,\xi_a$ take the place - in the bosonic case - of the Grassmannian 
variables $c^a,\bar{c}_a$, we simply rewrite the charges conserved in the {\it CPI} and replace in them
$c^a$ and $\bar{c}_a$ with $\pi^a$ and $\xi_a$. In the {\it CPI} the conserved charges are 
\cite{Gozzi}\cite{deotto}
\begin{equation}
\left\{
	\begin{array}{l}
	\displaystyle \label{4-11}
	Q_g=ic^a\bar{c}_a \;\;\;\;\;\;\;{\text{ghost charge}} \smallskip\\
	N=c^a\partial_aH \smallskip\\
	\overline{N}=\bar{c}_a\omega^{ab}\partial_bH
	\end{array}
	\right.
\end{equation}
and
\begin{equation}
\left\{
	\begin{array}{l}
	\displaystyle \label{4-12}
	Q=ic^a\lambda_a \;\;\;\;\;\;\;\;\;\;\;\;{\text{BRS charge}} \smallskip\\
	\overline{Q}=i\bar{c}_a\omega^{ab}\lambda_b \;\;\;\;\;\;\;{\text{anti-BRS charge}} \smallskip\\
	Q_{\scriptscriptstyle H}=Q-N \;\;\;\;\;\;\;{\text{susy-charge}} \smallskip\\
	\overline{Q}_{\scriptscriptstyle H}=\overline{Q}+\overline{N} \;\;\;\;\;\;\;{\text{susy-charge}}.
	\end{array}
	\right.
\end{equation}
So by replacing naively $c^a$ with $\pi^a$ and $\bar{c}_a$ with $\xi_a$ we get the two following set of charges
\begin{equation}
\left\{
	\begin{array}{l}
	\displaystyle \label{4-13}
	Q_g^{\scriptscriptstyle (B)}=i\pi^a\xi_a \smallskip\\
	N^{\scriptscriptstyle (B)}=\pi^a\partial_aH \smallskip\\
	\overline{N}^{\scriptscriptstyle (B)}=\xi_a\omega^{ab}\partial_bH
	\end{array}
	\right.
\end{equation}
and 
\begin{equation}
\left\{
	\begin{array}{l}
	\displaystyle \label{4-14}
	Q^{\scriptscriptstyle (B)}=i\pi^a\lambda_a \smallskip\\
	\overline{Q}^{\scriptscriptstyle (B)}=i\xi_a\omega^{ab}\lambda_b \smallskip\\
	Q_{\scriptscriptstyle H}^{\scriptscriptstyle (B)}=Q^{\scriptscriptstyle (B)}-N^{\scriptscriptstyle (B)} \smallskip\\
	\overline{Q}_{\scriptscriptstyle H}^{\scriptscriptstyle (B)}=\overline{Q}^{\scriptscriptstyle (B)}+\overline{N}^{\scriptscriptstyle (B)}
	\end{array}
	\right.
\end{equation}
where the superscript $(B)$ indicates that it refers to the {\it BFA} case. The reader may complain that by replacing 
$c^a$ with $\pi^a$ and $\bar{c}_a$ with $\xi_a$ we have not really done the replacement which would bring the 
$\widetilde{\cal H}$ of the {\it CPI} into the $\cal H$ of the {\it BFA} (\ref{3-5}). 
The difference is just in multiplicative 
factors $(\pm i)$ in front of the charges and this would not spoil their conservation. 
The careful reader may also notice that in the {\it CPI} there were other two other conserved charges \cite{Gozzi}
which are
\begin{equation}
\left\{
	\begin{array}{l}
	\displaystyle 
	K=\frac{1}{2}c^a\omega_{ab}c^b \smallskip\\
	\displaystyle \overline{K}=\frac{1}{2}\bar{c}_a\omega^{ab}\bar{c}_b.
	\end{array}
	\right.
\end{equation}
We did not list them because, via the substitution we did for the {\it BFA} case, the corresponding charges
would be zero because of the bosonic character of $\pi$ and $\xi$ and the antisymmetry of $\omega_{ab}$:
\begin{equation}
\left\{
	\begin{array}{l}
	\displaystyle 
	K^{\scriptscriptstyle (B)}=\frac{1}{2}\pi^a\omega_{ab}\pi^b=0 \smallskip\\
	\displaystyle \overline{K}^{\scriptscriptstyle (B)}=\frac{1}{2}\xi_a\omega^{ab}\xi_b=0.
	\end{array}
	\right.
\end{equation}
Turning now back to the set of charges in (\ref{4-13}), it is easy to check that they are all conserved, i.e.:
\begin{equation}
[Q_g^{\scriptscriptstyle (B)},\widehat{\cal H}_{\scriptscriptstyle BFA}]=
[N^{\scriptscriptstyle (B)},\widehat{\cal H}_{\scriptscriptstyle BFA}]=
[\overline{N}^{\scriptscriptstyle (B)},\widehat{\cal H}_{\scriptscriptstyle BFA}]=0. \label{4-15}
\end{equation}
The detailed calculations are given in Appendix C. On the other hand
the charges present in the second set given in (\ref{4-14}) are apparently not conserved. In fact taking the 
bosonic analog of the BRS charge and its commutator with $\widehat{\cal H}_{\scriptscriptstyle BFA}$ we get: 
\begin{equation}
[Q^{\scriptscriptstyle (B)},\widehat{\cal H}_{\scriptscriptstyle BFA}]=
-\pi^l\omega^{ab}\partial_b\partial_l\partial_cH\pi^c\xi_a \label{4-16}
\end{equation}
and for the anti-BRS charge 
\begin{equation}
[\overline{Q}^{\scriptscriptstyle (B)},\widehat{\cal H}_{\scriptscriptstyle BFA}]=
-\xi_a\omega^{ab}\xi_s\omega^{st}(\partial_b\partial_t\partial_lH)\pi^l.
\label{4-17}
\end{equation}
These straightforward calculations are reported in Appendix C. The charges we have not yet analyzed are the bosonic 
analogs of the supersymmetry charges $Q_{\scriptscriptstyle H}^{\scriptscriptstyle(B)},
\overline{Q}_{\scriptscriptstyle H}^{\scriptscriptstyle(B)}$ in (\ref{4-14}). They cannot be conserved because 
they are a linear combination of $Q^{\scriptscriptstyle (B)}$ and $\overline{Q}^{\scriptscriptstyle(B)}$,
which are not conserved, with $N^{\scriptscriptstyle (B)}$ and $\overline{N}^{\scriptscriptstyle (B)}$
which are conserved. 

Let us now turn to Eqs. (\ref{4-16}) and (\ref{4-17}) and let us look at their RHS which indicate by ``{\it how much}" the
conservation law is violated. It is easy to notice that these RHS do not contain $\lambda_a$ and so they commute
with the original phase space operators $\varphi^a$.
As a consequence of this we have that the infinitesimal transformations generated by $Q^{\scriptscriptstyle (B)}$
and by the Hamiltonian $\widehat{\cal H}_{\scriptscriptstyle BFA}$ {\it commute} when they are applied on $\varphi$. 
In fact the infinitesimal BRS transformation generated by $Q^{\scriptscriptstyle (B)}$ on a field $A$ is given
by the commutator of $Q^{\scriptscriptstyle (B)}$ with the field: $\delta_{Q^{\scriptscriptstyle (B)}}A=
[\epsilon Q^{\scriptscriptstyle (B)}, A]$ where $\epsilon$ is an infinitesimal parameter. 
The same happens for the transformations generated by the Hamiltonian:
$\delta_{\cal H}A=[\bar{\epsilon}\,\widehat{\cal H}_{\scriptscriptstyle BFA},A]$. 
Suppose we take for $A$ the original phase space variables
$\varphi^a$. If we perform first an infinitesimal time evolution and then a BRS transformation we obtain
\begin{equation}
\delta_{Q^{\scriptscriptstyle (B)}}\delta_{\cal H}\varphi^a=\epsilon\bar{\epsilon}\bigl[Q^{\scriptscriptstyle (B)}, 
[\widehat{\cal H}_{\scriptscriptstyle BFA},\varphi^a]\bigr]
\end{equation} 
while, if we perform the transformations in the inverse order, we obtain:
\begin{equation}
\delta_{\cal H}\delta_{Q^{\scriptscriptstyle (B)}}\varphi^a=
\epsilon\bar{\epsilon}\bigl[\widehat{\cal H}_{\scriptscriptstyle BFA}, [Q^{\scriptscriptstyle (B)},\varphi^a]\bigr].
\end{equation} 
Now we can use the Jacobi identities to obtain
\begin{eqnarray}
\delta_{Q^{\scriptscriptstyle (B)}}\delta_{\cal H}\varphi^a-
\delta_{\cal H}\delta_{Q^{\scriptscriptstyle (B)}}
\varphi^a&=&\epsilon\bar{\epsilon}\biggl(\bigl[Q^{\scriptscriptstyle (B)},[\widehat{\cal H}_{\scriptscriptstyle BFA},
\varphi^a]\bigr]-\bigl[\widehat{\cal H}_{\scriptscriptstyle BFA},[Q^{\scriptscriptstyle (B)},\varphi^a]\bigr]\biggr)=\nonumber\\
&=&-\epsilon\bar{\epsilon}\bigl[\varphi^a,[Q^{\scriptscriptstyle (B)},\widehat{\cal H}_{\scriptscriptstyle BFA}]\bigr]=0 
\label{zip}
\end{eqnarray}
where in the last step we have used the fact that the RHS of (\ref{4-16}) commutes with $\varphi^a$. 
``Somehow" we can say that the transformations generated by 
$Q^{\scriptscriptstyle (B)}$ and $\widehat{\cal H}_{\scriptscriptstyle BFA}$ 
commute on the original phase space. Of course
the same will happen for the anti-BRS charge $\overline{Q}^{\scriptscriptstyle (B)}$ and for the 
supersymmetry charges $Q_{\scriptscriptstyle H}^{\scriptscriptstyle (B)},
\overline{Q}_{\scriptscriptstyle H}^{\scriptscriptstyle (B)}$. 

Now we try to provide a geometrical
interpretation of this fact at least for the BRS-charge. Let us do an infinitesimal BRS transformation on $\varphi^a$:
\begin{equation}
\delta_{Q^{\scriptscriptstyle (B)}}\varphi^a=[\epsilon Q^{\scriptscriptstyle (B)},\varphi^a]=
[\epsilon i\pi^b\lambda_b,\varphi^a]=\epsilon\pi^a
\label{4-20}
\end{equation}
where $\epsilon$ is an infinitesimal commuting parameter. The new phase space point $\varphi^{\prime a}$ reached after 
this transformation is:
\begin{equation}
\varphi^{\prime a}=\varphi^a+\epsilon\pi^a. \label{4-21}
\end{equation}
Remember now that $\pi^a$ is a Jacobi field that means it satisfies the equation of the first variation
(\ref{3-8-b}). So if $\varphi^a$ is a point on a trajectory, $\varphi^{\prime a}$ is a point on a nearby trajectory
as indicated in the Figure 1 below:

\begin{center}
\begin{picture}(240,120)
\pscurve[]{c-c}(-0.9,2.4)(0.9,2.9)(2.9,3.2)(4.9,3.4)(6.9,3.5)
\pscurve[]{c-c}(-0.9,0.4)(0.9,0.7)(2.9,0.9)(4.9,1)(6.9,1.1)
\psline[]{->}(0.9,0.7)(0.9,1.8)
\psline[](0.9,1.8)(0.9,2.9)

\rput(0.9,3.4){$\varphi^{\prime a}=\varphi^a+\epsilon\pi^a$}
\rput(0.9,0.4){$\varphi^a$}
\rput(8.3,2.3){[Fig. 1]}
\end{picture}
\end{center}

From Fig. 1 we expect that we could move from the point $\varphi^a$ along its trajectory via the Hamiltonian 
$\widehat{\cal H}_{\scriptscriptstyle BFA}$ for an interval of time $\Delta t$, reach a point $\varphi_{\scriptscriptstyle (1)}^a$ and 
from there jump, via a
BRS transformation to a point $\varphi^{\prime a}_{\scriptscriptstyle(1)}$ on the nearby trajectory. 
Moving then back on this 
second trajectory for an interval of time $\Delta t$ we should reach the point $\varphi^{\prime a}$ that we originally
reached via a simple BRS transformation from $\varphi^a$. All this is illustrated in Fig. 2 below. 

\begin{center}
\begin{picture}(240,120)
\pscurve[]{c-c}(-0.9,2.4)(0.9,2.9)(2.9,3.2)(4.9,3.4)(6.9,3.5)
\pscurve[]{c-c}(-0.9,0.4)(0.9,0.7)(2.9,0.9)(4.9,1)(6.9,1.1)
\psline[](0.9,0.7)(0.9,1.8)
\psline[](0.9,1.8)(0.9,2.9)
\psline[](0.9,1.8)(0.8,1.7)
\psline[](0.9,1.8)(1.0,1.7)
\psline[](4.9,1)(4.9,2.2)
\psline[](4.9,2.2)(4.9,3.4)
\psline[](2.9,3.2)(3.0,3.1)
\psline[](2.9,3.2)(3.0,3.3)
\psline[](4.9,2.2)(4.8,2.1)
\psline[](4.9,2.2)(5.0,2.1)
\psline[](2.9,0.9)(2.8,1.0)
\psline[](2.9,0.9)(2.8,0.8)
\rput(1.4,1.8){\small{BRS}}
\rput(2.9,2.9){\small{$\widehat{\cal H}\Delta t$}}
\rput(2.9,0.5){\small{$\widehat{\cal H}\Delta t$}}

\rput(0.9,3.4){$\varphi^{\prime a}=\varphi^a+\epsilon\pi^a$}
\rput(0.9,0.4){$\varphi^a$}
\rput(4.9,3.8){$\varphi_{\scriptscriptstyle (1)}^{\prime a}=\varphi_{\scriptscriptstyle (1)}^a+\epsilon\pi^a$}
\rput(4.9,0.7){$\varphi_{\scriptscriptstyle (1)}^{a}$}
\rput(5.4,2.2){\small{BRS}}
\rput(8.3,2.3){[Fig. 2]}
\end{picture}
\end{center}

This diagram expresses the fact that, {\it in the} $\varphi$-{\it space}, 
the BRS transformation and $\widehat{\cal H}_{\scriptscriptstyle BFA}$ should commute
that is what Eq. (\ref{zip}) tells us. 

Let us now turn to the bosonic analogs of the susy charges $Q_{\scriptscriptstyle H}^{\scriptscriptstyle (B)},
\overline{Q}_{\scriptscriptstyle H}^{\scriptscriptstyle (B)}$. As they are linear combination
of $Q^{\scriptscriptstyle (B)},\overline {Q}^{\scriptscriptstyle (B)},N,\overline{N}$ and these last two charges commute 
with $\widehat{\cal H}_{\scriptscriptstyle BFA}$, from (\ref{4-16}) and (\ref{4-17}) we will get 
\begin{eqnarray}
&&[Q_{\scriptscriptstyle H}^{\scriptscriptstyle (B)},\widehat{\cal H}_{\scriptscriptstyle BFA}]=
-\pi^l\omega^{ab}\partial_b\partial_l\partial_cH\pi^c\xi_a\nonumber\\
&&[\overline{Q}_{\scriptscriptstyle H}^{\scriptscriptstyle (B)},\widehat{\cal H}_{\scriptscriptstyle BFA}]=
-\xi_a\omega^{ab}\xi_s\omega^{st}(\partial_b\partial_t\partial_lH)\pi^l.
\label{4-22}
\end{eqnarray}
So also the transformations generated by $Q_{\scriptscriptstyle H}^{\scriptscriptstyle (B)}$ and 
$\overline{Q}_{\scriptscriptstyle H}^{\scriptscriptstyle (B)}$
commute with those generated by $\widehat{\cal H}_{\scriptscriptstyle BFA}$ on the phase space variables $\varphi^a$.
It would be interesting to check whether they behave as true supersymmetry charges that means
\begin{equation}
[Q_{\scriptscriptstyle H}^{\scriptscriptstyle (B)},\overline{Q}^{\scriptscriptstyle (B)}_{\scriptscriptstyle H}]
=2\widehat{\cal H}_{\scriptscriptstyle BFA}.
\end{equation}
It is actually easy to work out the commutators of $Q_{\scriptscriptstyle H}^{\scriptscriptstyle (B)},
\overline{Q}_{\scriptscriptstyle H}^{\scriptscriptstyle (B)}$ and the calculation, presented in detail
in Appendix C, gives: 
\begin{equation}
[Q_{\scriptscriptstyle H}^{\scriptscriptstyle (B)},\overline{Q}_{\scriptscriptstyle H}^{\scriptscriptstyle (B)}]=
2\widehat{\cal H}_{\scriptscriptstyle BFA}+\{4\pi^a\omega^{de}\partial_e\partial_aH\xi_d\}. \label{4-23}
\end{equation}
We see that we can get the standard supersymmetry algebra if the last term on the RHS of (\ref{4-23}) were zero. Again this 
term does not contain $\lambda_a$ and so on the $\varphi^a$ variables we have that
\begin{equation}
\delta_{Q_{\scriptscriptstyle H}^{\scriptscriptstyle (B)}}
\delta_{\overline{Q}_{\scriptscriptstyle H}^{\scriptscriptstyle (B)}}
\varphi^a-\delta_{\overline{Q}_{\scriptscriptstyle H}^{\scriptscriptstyle (B)}}
\delta_{Q_{\scriptscriptstyle H}^{\scriptscriptstyle (B)}}\varphi^a=2\delta_{\widehat{\cal H}}\varphi^a.
\end{equation}
i.e. the supersymmetry algebra holds. 

Usually supersymmetry is described as the ``{\it square root}" of the time translation. Let us find out whether 
there is anything
like that in our bosonic case. Instead of the two charges $Q_{\scriptscriptstyle H}^{\scriptscriptstyle (B)}$ 
and $\overline{Q}_{\scriptscriptstyle H}^{\scriptscriptstyle (B)}$, let us build the following
other ones
\begin{equation}
\left\{
	\begin{array}{l}
	\displaystyle 
	Q_{1}^{\scriptscriptstyle (B)}=Q^{\scriptscriptstyle (B)}-\overline{N}^{\scriptscriptstyle (B)} \smallskip\\
	Q_{2}^{\scriptscriptstyle (B)}=\overline{Q}^{\scriptscriptstyle (B)}+N^{\scriptscriptstyle (B)}.
	\label{4-24}
	\end{array}
	\right.
\end{equation}
The transformations of our variables under $Q_{1}^{\scriptscriptstyle (B)}$ can be easily worked, 
and it is done in detail in Appendix C. The result is:
\begin{equation}
\left\{
	\begin{array}{l}
	\displaystyle 
	\delta_{Q_1^{\scriptscriptstyle (B)}}\varphi^a=\epsilon\pi^a \smallskip\\
	\delta_{Q_1^{\scriptscriptstyle (B)}}\xi_a=\epsilon\lambda_a \smallskip\\
	\delta_{Q_1^{\scriptscriptstyle (B)}}\pi^a=-i\epsilon\omega^{ae}\partial_eH \smallskip\\
	\delta_{Q_1^{\scriptscriptstyle (B)}}\lambda_a=-i\epsilon \xi_b\omega^{be}(\partial_e\partial_aH)
	\label{4-25}
	\end{array}
	\right.
\end{equation}
where $\epsilon$ is an infinitesimal commuting parameter. Let us check whether, by doing these transformations twice, 
we get a time translation. Let us just do it on the original phase space $\varphi^a$. Using (\ref{4-25}) we get
\begin{equation}
\delta^2_{Q_1^{\scriptscriptstyle (B)}}\varphi^a=\delta_{Q_1^{\scriptscriptstyle (B)}}(\epsilon\pi^a)=
-i\epsilon^2\omega^{ae}\partial_eH=-i\epsilon^2\dot{
\varphi}^a. \label{4-26}
\end{equation}
In the last step above we have used the equations of motion. The result seems to confirm that 
$Q_1^{\scriptscriptstyle (B)}$ is the ``{\it
square root}" of the time translation. Eq. (\ref{4-26}) is an infinitesimal time translation if we equate
$\epsilon^2=\Delta t$. So we could say that, in order
to do an infinitesimal time translation, we could perform two $Q_1^{\scriptscriptstyle (B)}$ transformations in a row each with 
``infinitesimal" parameter $\epsilon$ related to the ``square root" of $\Delta t$. 
We find that it is curious that, at least on some hypersurfaces of our $8n$-dimensional space we could,
without introducing Grassmannian variables and
via purely bosonic charges, get something analogous to supersymmetry or better to the square
root of a time translation. 

Let us now go back to geometry and to the bosonic BRS charge $Q^{\scriptscriptstyle (B)}$. 
In the Grassmannian
case the BRS charge can be identified \cite{Gozzi} with the exterior derivative. One of the properties \cite{marsden}
of the exterior derivative is that it commutes with the Lie derivative. This is not anymore the case for our
$Q^{\scriptscriptstyle (B)}$ as it is proved in (\ref{4-16}). Even if it does so in the $\varphi$-space, it is not enough. In fact the exterior
derivative must commute with the Lie derivative in the full space of forms which is somehow an extension
of the ordinary phase space. Actually it is the space of higher forms which has to be properly defined in 
the {\it BFA} case and this is
what we will do in the next section. 

\section{Higher Forms}

We listed in Eq. (\ref{4-4}) which variables to use in order to build one-forms. They are the operators $\widehat{\pi}^a$ which take
the place, in the bosonic case, of the Grassmannian variables $c^a$. The problem arises when we want to build higher forms. 
We know that between forms one defines the so called {\it wedge} product $\wedge$ so that, for example, the basis
$d\varphi^a\wedge d\varphi^b$ for two-forms is antisymmetric in the interchange of $a\leftrightarrow b$. This was naturally
incorporated in the Grassmannian formalism \cite{Gozzi} by representing the forms $d\varphi^a$ with Grassmannian variables
$c^a$. They are anticommuting and so the antisymmetry of $d\varphi^a\wedge d\varphi^b$ is 
automatically produced by the antisymmetry 
of the product $c^ac^b$:
\begin{equation}
	\begin{array}{c}
	\displaystyle 
	d\varphi^a\wedge d\varphi^b=-d\varphi^b\wedge d\varphi^a \smallskip\\
	\Updownarrow \smallskip\\
	c^ac^b=-c^bc^a.
	\label{5-1}
	\end{array}
\end{equation}
In the bosonic case we do not have Grassmannian variables and the forms $\widehat{\pi}^a$ commute among themselves so that, 
if we represent a two-form $d\varphi^a\wedge d\varphi^b$ as $\widehat{\pi}^a\widehat{\pi}^b$, 
we loose its anticommuting nature.

The way out seems to be the standard procedure used in the literature \cite{eguchi} on differential geometry, i.e.
to introduce a tensor product among the cotangent spaces whose basis are the $d\varphi^a$ and define the wedge product 
$\wedge$ as
\begin{equation}
\displaystyle 
\label{tensorpr}
d\varphi^a\wedge d\varphi^b\equiv \frac{1}{2}(d\varphi^a\otimes d\varphi^b-d\varphi^b\otimes d\varphi^a).
\end{equation}
In our case the role of the $d\varphi^a$ is taken by the operator $\widehat{\pi}^a$ and we should build tensor 
products among them. To do that we have to enlarge our Hilbert space. Originally it was made of functions 
$\psi(\varphi^a,\pi^a)$ which could be considered as belonging to the tensor product of the two Hilbert spaces of 
the wave functions $\psi(\varphi)$ and $\widetilde{\psi}(\pi)$  
which we indicate as:
\begin{equation}
{\mathscr H}\equiv {\mathscr H}_{\varphi}\otimes {\mathscr H}_{\pi}. \label{5-2}
\end{equation}
From now on the new Hilbert space we will use is the following one 
\begin{equation}
{\mathscr H}_{2n}\equiv {\mathscr H}_{\varphi}\otimes {\mathscr H}_{\pi_{(1)}}
\otimes {\mathscr H}_{\pi_{(2)}}\ldots\otimes {\mathscr H}_{\pi_{(2n)}}. \label{5-3}
\end{equation}
where we have made the tensor products of copies of the space ${\mathscr H}_{\pi}$ and labeled them 
with ${\mathscr H}_{\pi_{(1)}}$, $\ldots {\mathscr H}_{\pi_{(n)}}$.
If we limit ourselves to the case $n=1$ the space (\ref{5-3}) reduces to: 
\begin{equation}
{\mathscr H}_2\equiv {\mathscr H}_{\varphi}\otimes {\mathscr H}_{\pi_{(1)}}
\otimes {\mathscr H}_{\pi_{(2)}} \label{5-5}
\end{equation}
and in this space we have that, for example, a two-form is represented as 
\begin{equation}
\displaystyle \widehat{F}
=F_{ab}(\widehat{\varphi})\otimes \frac{1}{2}[\widehat{\pi}_{\scriptscriptstyle (1)}^a\otimes
\widehat{\pi}^b_{\scriptscriptstyle (2)}-\widehat{\pi}_{\scriptscriptstyle (1)}^b\otimes
\widehat{\pi}^a_{\scriptscriptstyle (2)}]. \label{5-4}
\end{equation}
The operator we have used
up to now to represent the Lie derivative, which is (\ref{3-5}), was a good one but only for the space (\ref{5-2}).
The new space is (\ref{5-5}) and we should extend 
$\widehat{\cal H}_{\scriptscriptstyle BFA}$ of (\ref{3-5}) to this space. 
We could try this operator
\begin{equation}
\widehat{\cal H}\equiv \widehat{\lambda}_a\omega^{ab}\partial_bH(\widehat{\varphi})\otimes 
{\bf 1}_{\scriptscriptstyle (1)}\otimes {\bf 1}_{\scriptscriptstyle (2)}
-\omega^{be}\partial_e\partial_a H(\widehat{\varphi})\otimes (\widehat{\pi}^a_{\scriptscriptstyle (1)}
\widehat{\xi}_b^{\scriptscriptstyle (1)}\otimes {\bf 1}_{\scriptscriptstyle (2)}
+{\bf 1}_{\scriptscriptstyle (1)}\otimes 
\widehat{\pi}^a_{\scriptscriptstyle (2)}\widehat{\xi}_b^{\scriptscriptstyle (2)}). \label{5-6}
\end{equation}
Using as commutators the following ones 
\begin{eqnarray}
&&[\widehat{\pi}_{\scriptscriptstyle (i)}^a,\widehat{\pi}_{\scriptscriptstyle (j)}^b]=0\nonumber\\
&&[\widehat{\xi}^{\scriptscriptstyle (i)}_a,\widehat{\xi}^{\scriptscriptstyle (j)}_b]=0\nonumber\\
&&[\widehat{\xi}_a^{\scriptscriptstyle (i)},\widehat{\pi}_{\scriptscriptstyle (j)}^b]=
i\delta_a^b\delta^{\scriptscriptstyle (i)}_{\scriptscriptstyle (j)} \label{5-7}\\
&&[\widehat{\varphi}^a,\widehat{\pi}_{\scriptscriptstyle (i)}^b]=0 \nonumber\\
&&[\widehat{\varphi}^a,\widehat{\xi}_b^{\scriptscriptstyle (i)}]=0 \nonumber
\end{eqnarray}
it is easy to see that the action of $\widehat{\cal H}$ presented in (\ref{5-6}) 
on the two-form $\widehat{F}$ of (\ref{5-4})
is
\begin{equation}
\displaystyle [i\widehat{\cal H},\widehat{F}]=
\omega^{ab}[\partial_bH\partial_aF_{de}+\partial_b\partial_dHF_{ae}
+\partial_b\partial_eHF_{da}]\otimes \frac{1}{2}(\widehat{\pi}^d_{\scriptscriptstyle (1)}\otimes
\widehat{\pi}^e_{\scriptscriptstyle (2)}-
\widehat{\pi}^e_{\scriptscriptstyle (1)}
\otimes\widehat{\pi}^d_{\scriptscriptstyle (2)}) \label{5-8}
\end{equation}
and this is exactly the manner how two-forms transform under the Lie derivative \cite{marsden}. The derivation of 
(\ref{5-8}) is presented in detail in Appendix D. Eq. (\ref{5-8}) confirms that the choice 
of $\widehat{\cal H}$ we made in (\ref{5-6}) is the correct one. 
In the case $n=1$ we have only zero-, one- and two-forms and we have already seen that two-forms are represented
by Eq. (\ref{5-4}). How are zero- and one-forms represented? The zero-forms $\widehat{F}$
and the one-forms $\widehat{C}$ are respectively
\begin{equation}
\widehat{F}
=F(\widehat{\varphi})\otimes [{\bf 1}_{\scriptscriptstyle (1)}\otimes {\bf 1}_{\scriptscriptstyle (2)}],
\end{equation}
\begin{equation}
\widehat{C}
=C_d(\widehat{\varphi})\otimes [\widehat{\pi}_{\scriptscriptstyle (1)}^d\otimes {\bf 1}_{\scriptscriptstyle (2)}+
{\bf 1}_{\scriptscriptstyle (1)}\otimes \widehat{\pi}^d_{\scriptscriptstyle (2)}]. \label{oneforms}
\end{equation}
It should be clear by now that the indices $(1)$ and $(2)$ indicate the $\widehat{\mathscr H}_{\pi_{\scriptscriptstyle (1)}}$
and $\widehat{\mathscr H}_{\pi_{\scriptscriptstyle (2)}}$ appearing in (\ref{5-5}).
As we will prove in detail in Appendix D, 
the commutator of $i\widehat{\cal H}$ with $\widehat{C}$
gives the correct action of the Lie derivative on one-forms:
\begin{equation}
[i\widehat{\cal H},\widehat{C}]=
(\partial_aC_d\omega^{ab}\partial_bH+\omega^{ae}\partial_e\partial_dHC_a)
\otimes[\widehat{\pi}^d_{\scriptscriptstyle (1)}\otimes{\bf 1}_{\scriptscriptstyle (2)}
+{\bf 1}_{\scriptscriptstyle (1)}
\otimes\widehat{\pi}^d_{\scriptscriptstyle (2)}]. \label{comm1}
\end{equation}
So we can conclude that, in the case $n=1$, the operator (\ref{5-6}) represents a good extension of the Lie derivative on 
the entire space of differential forms.

It is easy to generalize 
$\widehat{\cal H}$ of (\ref{5-6})
to the Lie derivative which acts in a space with an arbitrary number $n$ of degrees of freedom. 
It is the following one
\begin{equation}
\widehat{\cal H}\equiv\lambda_a\omega^{ab}\partial_bH\otimes{\bf 1}^{\otimes 2n}-\omega^{be}\partial_e\partial_aH\otimes
{\bf S}[\widehat{\pi}^a\widehat{\xi}_b\otimes{\bf 1}^{\otimes (2n-1)}] \label{5-9}
\end{equation}
where by ${\bf 1}^{\otimes 2n}$ we indicate the tensor product of $2n$ identity operators, and with ${\bf S}$ the 
symmetrization operation of the operators contained in the square brackets. So for example for $n=2$ we have
\begin{eqnarray}
{\bf S}[\widehat{\pi}^a\widehat{\xi}_b\otimes{\bf 1}^{\otimes 3}]&=&
\widehat{\pi}_{\scriptscriptstyle (1)}^a\widehat{\xi}_b^{\scriptscriptstyle (1)}
\otimes {\bf 1}_{\scriptscriptstyle (2)}\otimes {\bf 1}_{\scriptscriptstyle (3)}
\otimes {\bf 1}_{\scriptscriptstyle (4)}+{\bf 1}_{\scriptscriptstyle (1)} \otimes 
\widehat{\pi}^a_{\scriptscriptstyle (2)}\widehat{\xi}_b^{\scriptscriptstyle (2)}
\otimes {\bf 1}_{\scriptscriptstyle (3)}\otimes {\bf 1}_{\scriptscriptstyle (4)}+\nonumber\\
&&+{\bf 1}_{\scriptscriptstyle (1)}\otimes{\bf 1}_{\scriptscriptstyle (2)}
\otimes\widehat{\pi}^a_{\scriptscriptstyle (3)}\widehat{\xi}_b^{\scriptscriptstyle (3)}
\otimes{\bf 1}_{\scriptscriptstyle (4)}+{\bf 1}_{\scriptscriptstyle (1)}\otimes {\bf 1}_{\scriptscriptstyle (2)}\otimes
{\bf 1}_{\scriptscriptstyle (3)}\otimes 
\widehat{\pi}^a_{\scriptscriptstyle (4)}\widehat{\xi}_b^{\scriptscriptstyle (4)}. \label{5-10}
\end{eqnarray}
Let us always remember that the indices $(1),(2),\ldots (2n)$ indicate to which Hilbert space
${\mathscr H}_{\pi_{(i)}}$ in (\ref{5-3}) the operators $\widehat{\pi}_{\scriptscriptstyle (i)}$,
$\widehat{\xi}^{\scriptscriptstyle (i)}$ and ${\bf 1}_{\scriptscriptstyle (i)}$ belong.
In the same way it is possible to generalize the concept of differential form. An $m$-form in a $2n$-dimensional space
is given by:
\begin{equation}
\displaystyle \widehat{P}\equiv{\bf S}\biggl\{\frac{1}{m!}P_{a_1\ldots a_m}(\widehat{\varphi})
\otimes
{\bf A}\{\widehat{\pi}_{\scriptscriptstyle (1)}^{a_1}\otimes\widehat{\pi}_{\scriptscriptstyle (2)}^{a_2}\otimes\ldots
\widehat{\pi}_{\scriptscriptstyle (m)}^{a_m}\}\otimes{\bf 1}^{\otimes (2n-m)}\biggr\}, \label{genform}
\end{equation}
where ${\bf A}$ indicates the antisymmetrizer of the basis $\widehat{\pi}_{\scriptscriptstyle (i)}^{a_i}$
of the $m$ cotangent spaces needed to build an $m$-form. The position of this $m$ operators $\widehat{\pi}$
inside the string of the $2n$ Hilbert spaces is completely arbitrary. Therefore if we do not want to choose
a particular position we can symmetrize the $2n-m$ identity operators with the $m$ operators $\widehat{\pi}$
by means of the symmetrizer ${\bf S}$ as we did in Eq. (\ref{oneforms}) for the one-forms.
As we will prove in detail in Appendix D, the commutator
$[i\widehat{\cal H},\widehat{P}]$ reproduces the correct action of the Lie derivative on an arbitrary differential form 
$P$:
\begin{equation}
[i\widehat{\cal H},\widehat{P}]={\cal L}_{(dH)^{\sharp}}
P. \label{genlie}
\end{equation} 
 
Besides the forms we can build, using the variables
$\widehat{\xi}_a$, the symmetric tensors. This last operation was not possible 
in the Grassmannian or {\it CPI} formalism \cite{Gozzi}. We will hopefully come back to these issues in the future.

Before concluding this section let us notice that, differently than in the {\it CPI} case, the higher forms are not
represented by {\it wave functions} $\psi(\varphi,c)$ of the theory but by {\it operators} like in
(\ref{genform}). In fact wave functions, in the
bosonic case, would be generic functions $\psi(\varphi,\pi)$ and they would not have the structure which Grassmannian
ones do have:
\begin{equation}
\psi(\varphi,c)=\psi_0(\varphi)+\psi_a(\varphi)c^a+\psi_{ab}c^ac^b+\ldots +\psi_{abc\ldots l}c^ac^bc^c\ldots c^l.
\end{equation}
It was this structure which allowed us  to identify $\psi_0(\varphi)$ with zero-forms, $\psi_a(\varphi)c^a$ with one-forms, 
$\psi_{ab}c^ac^b$ with 
two-forms etc. In general in the bosonic case $\psi(\varphi,\pi)$ is a generic function of $\pi$ and this 
forbids the identification with forms. Moreover, as we said previously, a one-form would be represented by 
$\psi(\varphi,\pi)=\psi_a\pi^a$ which would be not an acceptable wave function because it is not normalizable. 
Of course this does not mean that in the formalism given by (\ref{5-9}) 
we cannot introduce wave functions. We can but they do not 
have the meaning of higher forms. Only operators like (\ref{genform}) have this meaning. 

The wave functions associated to the multi-form formalism of the Hamiltonian (\ref{5-9}) are basically those which make up
the Hilbert space (\ref{5-3}) and they are $\psi(\varphi^a,\pi^a_{\scriptscriptstyle (1)},
\pi^a_{\scriptscriptstyle (2)},\ldots,\pi^a_{\scriptscriptstyle (2n)})$. It is possible
to introduce also in this space a positive definite scalar product like we did in (\ref{3-3}) for the space (\ref{5-2}).
The result is
\begin{equation}
\langle \psi_1|\psi_2\rangle\equiv \int d^{2n}\varphi^a\prod_{i=1}^{2n}d^{2n}\pi_{\scriptscriptstyle
(i)}^a\,\psi_1^*\psi_2
\label{5-11}
\end{equation}
and it is easy to prove that with this product 
the Hamiltonian (\ref{5-9}) is Hermitian (see Appendix E for further details). 

The reader may remember that our original $\widehat{\cal H}_{\scriptscriptstyle BFA}$ (\ref{3-5}) was derived from the path integral formalism
(\ref{2-5-a}). A
natural question to ask is if also the multi-form Hamiltonian (\ref{5-9}) can be derived from a path integral. The answer 
is yes and the path integral is the following one
\begin{equation}
\displaystyle 
Z=\int {\mathscr D}\varphi^a{\mathscr D}\lambda_a\prod_{i=1}^{2n}{\mathscr D}\pi^a_{\scriptscriptstyle (i)}{\mathscr D}
\xi_a^{\scriptscriptstyle (i)} \,e^{i\int dt \,{\cal L}_{\scriptscriptstyle{MF}}}
\label{5-12-a}
\end{equation}
where the multiform (MF) Lagrangian ${\cal L}_{\scriptscriptstyle{MF}}$ is  
\begin{equation}
\displaystyle 
{\cal L}_{\scriptscriptstyle{MF}}=
\lambda_a\dot{\varphi}^a+\sum_{i=1}^{2n}\pi^a_{\scriptscriptstyle (i)}\dot{\xi}_a^{\scriptscriptstyle (i)}
-{\cal H}_{\scriptscriptstyle{MF}}
\end{equation}
with
\begin{equation}
{\cal H}_{\scriptscriptstyle{MF}}=
\lambda_a\omega^{ab}\partial_bH-\sum_{i=1}^{2n}\pi^a_{\scriptscriptstyle (i)}
\omega^{be}\partial_e\partial_aH\xi_b^{\scriptscriptstyle (i)}.
\label{5-12}
\end{equation}
At this level the proof of the hermiticity of ${\cal H}_{\scriptscriptstyle MF}$ under the scalar product 
(\ref{5-11}) is identical to the proof given in Eq. (\ref{3-6}). 
Basically in the Hamiltonian (\ref{5-12}) we have a set of extra variables $(\pi_{\scriptscriptstyle (i)},
\xi^{\scriptscriptstyle (i)})$
for each extra Hilbert space ${\mathscr H}_{\pi_{(i)}}$ appearing in (\ref{5-3}).
It is actually then easier to work with the Hamiltonian ${\cal H}_{\scriptscriptstyle{MF}}$ 
than with the one in (\ref{5-9}). We can in fact turn 
the $\pi_{\scriptscriptstyle (i)}^a, \xi_a^{\scriptscriptstyle (i)}$ into operators by just looking at the 
kinetic term of (\ref{5-12-a}) and the 
commutators we can derive from it are basically those that we introduced by hand in (\ref{5-7}) plus the usual one 
$[\varphi^a,\lambda_b]=i\delta_b^a$. This confirms that the path integral (\ref{5-12-a})
is basically the one behind the operatorial
formalism (\ref{5-9}). Unfortunately this path integral does not appear to have a ``natural" interpretation like 
the one in (\ref{2-5-a}) had, in the sense that the latter is naturally related to (\ref{2-2}) and (\ref{2-1}) which
are just Dirac deltas on the classical paths. These Dirac deltas are natural objects in a functional approach 
to {\it CM} because they just give weight one to classical paths and weight zero to non-classical ones. Nothing like that
can be done for the path integral (\ref{5-12-a}) which can be turned into a Dirac delta of the equations of motion
like in (\ref{2-2}) but it gets multiplied not by one determinant but by $2n$ of them. This structure does not allow
us to pass to Dirac deltas of the classical trajectories like we did in (\ref{2-1}). So somehow the path integral 
(\ref{5-12-a}) does not have a simple intuitive understanding. This is the price we pay: we have a formalism
with a positive definite scalar product and a Hermitian Hamiltonian but a physical understanding is lacking. 
If on the contrary we keep the intuitive single particle path integral associated to the Hilbert space (\ref{5-2})
then the tensor product structure $\otimes$, needed to build higher forms like in (\ref{genform}), 
has to be given from outside 
and is not provided directly by the path integral.
On the contrary in the Grassmannian or {\it CPI} case \cite{Gozzi} the whole formalism, even for higher forms, has 
a nice and intuitive 
understanding and construction (because it can be reduced to a Dirac delta on classical paths), and no extra structure
has to be brought in from outside, but the price
we paid is that we have to give up one of the two conditions: either the positive definiteness of the scalar
product or the hermiticity of the Hamiltonian. 

\section{Metaplectic Representation}

One of the crucial concepts we have used so far is that of Lie derivative \cite{marsden}. We have seen how it acts on 
vector fields (\ref{4-6}), on forms (\ref{4-8}) or on tensors in the case of {\it symplectic} manifolds. The notion of Lie
derivative can be generalized to {\it general} manifolds ${\cal M}_n$ whose group of diffeomorphism we indicate with 
$\text{Diff}({\cal M}_n)$ and whose {\it structure group} \cite{eguchi} of the associated\break
(co-)tangent bundle we indicate
with ${\cal G}$. An arbitrary tensor field ${\cal X}$ on ${\cal M}_n$ under the action of an element of $\text{Diff}({\cal
M}_n)$, which drags the field through an infinitesimal displacement $\delta\varphi^a=-h^a(\varphi)$, is transformed as
follows
\begin{equation}
{\cal X}^{\prime}(\varphi)-{\cal X}(\varphi)={\cal L}_h{\cal X}(\varphi) \label{6-1}
\end{equation}
where ${\cal L}_h$ is the Lie derivative associated to the vector field $h=h^a\partial_a$. The general abstract expression
\cite{dewitt} of ${\cal L}_h$ is:
\begin{equation}
{\cal L}_h=h^a\partial_a-\partial_bh^aG^b_{\; a} \label{6-2}
\end{equation}
where $G^b_{\;a}$ are the generators of the structure group ${\cal G}$ in the representation to which ${\cal X}$ belong.
The
indices $a,b$ are group indices and not representation indices which we shall indicate with $\alpha,\beta$. So the matrix
representation of $(G^a_{\;b})$ will be $(G^a_{\;b})^{\alpha}_{\;\beta}$ where $\alpha$ are also the indices of ${\cal X}$, if we
organize it as a vector. For a generic manifold we have that 
${\cal G}=\text{GL}(n,R)$, for a Riemann manifold ${\cal G}=O(n)$,
and for a symplectic manifold ${\cal G}=\text{Sp}(2N)$. It is easy to check that, if ${\cal X}$ is a vector or a form and 
${\cal M}_{2N}$ a symplectic manifold, expression (\ref{6-1}) reduces to the usual transformations (\ref{4-6}) and
(\ref{4-8}). This calculation is shown in Appendix F which, together with much of this section, is taken from Refs.
\cite{meta} and \cite{private}. As we said the coefficients $G^b_{\; a}$ are the generators of the structure group ${\cal G}$ in the
representation to which ${\cal X}$ belong. Now ${\cal G}$ could have also spinor representations like it is the case for
$O(n)$. Does this mean that we can introduce the concept of Lie derivative also for spinors besides vectors, forms and
tensors? The answer is yes but not along all vector fields $h^a$. In the Riemann case it is only along Killing vector
fields \cite{dewitt} and in the symplectic case only along Hamiltonian vector fields, which are those which preserve the
symplectic two-form
\begin{equation}
{\cal L}_h\omega=0
\end{equation}
and whose local expression is
\begin{equation}
h^a=\omega^{ab}\partial_bH \label{6-3}
\end{equation}
with $H(\varphi)$ a function on ${\cal M}_{2N}$. We will not give details of why we have to restrict ourselves to these
particular vector fields for spinor ${\cal X}$ but refer the reader to the literature \cite{dewitt}. Basically it is only
for those fields that the usual commutator structure of the Lie derivative is preserved even for spinors.

Before proceeding let us rewrite (\ref{6-2}) in a slightly modified form. Let us first introduce the following objects
\begin{eqnarray}
&& K_{ab}(\varphi)\equiv\partial_a\partial_bH(\varphi)\nonumber\\
&& \Sigma^{ab}\equiv i(\omega^{ca}G^b_{\;c}+\omega^{cb}G^a_{\;c}).
\end{eqnarray}
Both are symmetric in $a,b$ and
\begin{equation}
K_{ab}\Sigma^{ab}=2i\partial_bh^cG^b_{\; c}
\end{equation}
so (\ref{6-2}) can be rewritten as
\begin{equation}
{\cal L}_h=h^a\partial_a+\frac{i}{2}K_{ab}\Sigma^{ab}. \label{6-4}
\end{equation}
This is the classical Lie derivative for fields whose components transform, under infinitesimal $\text{Sp}(2N)$
transformations of the tangent space, via the operator
\begin{equation}
\displaystyle S=1-\frac{i}{2}K_{ab}\Sigma^{ab}. \label{6-5}
\end{equation}
This will be proved in Appendix F for forms and vectors. 
We want now to apply this formalism to spinors that means we want to use for $\Sigma^{ab}$ in (\ref{6-5}) the spinorial
representation of the $\text{Sp}(2N)$. To do that we have to pass to the universal covering group of $\text{Sp}(2N)$
which is the metaplectic group $\text{Mp}(2N)$ \cite{littlejohn}-\cite{konstant}-\cite{dewitt2}. In analogy to the
spinorial representation of the Lorentz group, we first have to build the representation of the Clifford algebra
\begin{equation}
\gamma^{\mu}\gamma^{\nu}+\gamma^{\nu}\gamma^{\mu}=2g^{\mu\nu} \label{6-6}
\end{equation}
which, in the case of the metaplectic group \cite{dewitt2}, is 
\begin{equation}
\gamma^a\gamma^b-\gamma^b\gamma^a=2i\omega^{ab}. \label{6-7}
\end{equation}
This algebra, because of the crucial minus sign difference on the LHS of (\ref{6-7}) with respect to (\ref{6-6}), does not
admit finite dimensional unitary representation. The reason is the same as the one for which we
cannot find any finite dimensional representation for the $\widehat{q},\widehat{p}$ in {\it QM}. They obey the algebra
$\widehat{q}\widehat{p}-\widehat{p}\widehat{q}=i\hbar$ and taking the trace on both sides, 
if they were represented by finite dimensional matrices, we would get
a contradictory result. The only representations are infinite dimensional. We will indicate this infinite dimensional
Hilbert space as ${\cal V}$ and with ``$x$" the indices of the vectors in this space. So the matrix $\gamma^a$ in 
(\ref{6-7}) will have an infinite matrix representation indicated by $(\gamma^a)^x_{\;y}$. The generators of the
metaplectic group will be operators in this Hilbert space. As the $\text{Mp}(2N)$ is the covering group of
$\text{Sp}(2N)$ there will be two elements $M(S)$ and $-M(S)$ of $\text{Mp}(2N)$ associated to each element $S\in
\text{Sp}(2N)$. Correspondingly the multiplication rules will be 
\begin{equation}
M(S_1)M(S_2)=\pm M(S_1S_2).
\end{equation}
Following the procedure used for spinors in the case of the Lorentz group, we now try to find an operator $M(S)$ on
${\cal V}$ such that 
\begin{equation}
M(S)^{-1}\gamma^aM(S)=S^a_{\;b}\gamma^b. \label{6-8}
\end{equation}
We choose $S^a_{\;b}$ infinitesimally close to the identity and parametrize it like in (\ref{6-5}). For $M(S)$ we 
make the ansatz
\begin{equation}
\displaystyle M(S)=1-\frac{i}{2}K_{ab}\Sigma^{ab}_{\text{meta}} \label{6-9}
\end{equation}
where with $\Sigma^{ab}_{\text{meta}}$ we indicate the operators $\Sigma^{ab}$ in the metaplectic representation. 
Let us now insert (\ref{6-9}) into (\ref {6-8}). The result \cite{meta} is
\begin{equation}
\displaystyle \Sigma_{\text{meta}}^{ab}=\frac{1}{4}(\gamma^a\gamma^b+\gamma^b\gamma^a). \label{6-9-a}
\end{equation}
So a representation of the matrix $\gamma^a$ gives rise to a corresponding representation of the generators
$\Sigma^{ab}_{\text{meta}}$. We consider only representations in which the $\gamma^a$ are Hermitian with respect 
to the inner product in ${\cal V}$. As a consequence also the $\Sigma_{\text{meta}}^{ab}$ is Hermitian and 
$M(S)$ turns out to be unitary:
\begin{eqnarray}
(\gamma^a)^{\dagger}=\gamma^a\nonumber\\
(\Sigma^{ab}_{\text{meta}})^{\dagger}=\Sigma^{ab}_{\text{meta}} \label{em}\\
M(S)^{\dagger}=M(S)^{-1}.\nonumber
\end{eqnarray}
Explicit representations of the symplectic operators 
$\gamma^a$ have been worked out and can be found in the literature \cite{meta}.
We will briefly review one of them here.
The Clifford algebra (\ref{6-7}) is isomorphic to the standard Heisenberg algebra made of $N$ positions $\widehat{x}^k$ and $N$
momenta operators $\widehat{p}^j$:
\begin{eqnarray}
&&[\widehat{x}^k,\widehat{p}^j]=i\hbar\delta^{kj} \label{6-10}\nonumber\\
&&[\widehat{x}^k,\widehat{x}^j]=0\\
&&[\widehat{p}^k,\widehat{p}^j]=0\nonumber\\
&&\qquad k,j=1,\ldots, N.\nonumber
\end{eqnarray}
Combining $\widehat{x}^k$ and $\widehat{p}^k$ into a single variable
\begin{equation}
\widehat{\phi}^a=(\widehat{p}^k,\widehat{x}^k) \qquad\qquad a=1,\ldots, 2N
\end{equation}
we get that the algebra (\ref{6-10}) can be written as:
\begin{equation}
\widehat{\phi}^a\widehat{\phi}^b-\widehat{\phi}^b\widehat{\phi}^a=i\hbar\omega^{ab} \label{comm}
\end{equation}
and this is isomorphic to the metaplectic analog of the Clifford algebra (\ref{6-7}); so we have the following 
representation for the $\gamma$:
\begin{equation}
\displaystyle \gamma^a=\biggl(\frac{2}{\hbar}\biggr)^{\frac{1}{2}}\widehat{\phi}^a.
\end{equation}
In the ``Schr\"odinger" representation in which $\widehat{x}^k$ is diagonal, we have
\begin{eqnarray}
&& (\gamma^k)^x_{\;y}=\biggl(\frac{2}{\hbar}\biggr)^{\frac{1}{2}}\langle x|\widehat{x}^k|y\rangle=
\biggl(\frac{2}{\hbar}\biggr)^{\frac{1}{2}}x^k\delta^N(x-y)\nonumber\\
&& (\gamma^{N+k})^x_{\;y}=\biggl(\frac{2}{\hbar}\biggr)^{\frac{1}{2}}\langle x|\widehat{p}^k|y\rangle=
-i(2\hbar)^{\frac{1}{2}}\partial_k\delta^N(x-y).
\end{eqnarray}
The indices $x,y$ are the Hilbert space indices we introduced before. 
With the representation above and using Eq.
(\ref{6-9-a}), we get the following expression for the $K_{ab}\Sigma^{ab}_{\text{meta}}$ entering the 
matrix $M(S)$ of
(\ref{6-9}):
\begin{eqnarray}
\displaystyle
\biggl(\frac{1}{2}K_{ab}\Sigma^{ab}_{\text{meta}}\biggr)^x_{\;y}&=&\biggl[-\frac{1}{2}K_{kj}\partial^k\partial^j
-\frac{1}{2}iK_{N+k,j}(x^k\partial^j+\partial^jx^k)+\nonumber\\
&&+\frac{1}{2}K_{N+k,N+j}x^kx^j\biggr]\delta^N(x-y) \label{6-11-b}
\end{eqnarray}
where we have put $\hbar=1$.

The geometrical picture we have so far is the following. We have a base space which is our phase space ${\cal M}_{2N}$
and on the fibers we have requested that the structure group $\text{Sp}(2N)$ no longer acts in the {\it vector}
representation like in Ref. \cite{Gozzi} or like in Secs. II-V of this paper with the
variables $\pi^a$, but in the {\it spinor} representation. The bundle we get is the analog of the ``spin-bundle" \cite{eguchi} but each 
fiber is
a Hilbert space ${\cal V}$. So we end up in a Hilbert bundle which we call ${\cal V}_{\varphi}$ to indicate that there
is a fiber ${\cal V}$ at each point $({\cal \varphi})$ of the phase space ${\cal M}_{2N}$. In each
fiber a state $|\psi\rangle$ can be represented in its basis $\langle x|$ by:
\begin{equation}
\psi^x\equiv\langle x|\psi\rangle \label{fiber}
\end{equation}
and we can introduce the dual state $\langle\psi|\in{\cal V}^*$ as
\begin{equation}
\langle\psi|x\rangle=(\psi^x)^*.
\end{equation}
The dual pairing is then the usual inner product
\begin{equation}
\langle {\cal X}|\psi\rangle\equiv\int d^Nx({\cal X}^x)^*\psi^x
\end{equation}
on $L^2(R^N,d^Nx)$. Note that the bra $\langle x|$ entering (\ref{fiber}) or the operators $\widehat{\phi}$ entering (\ref{comm})
have nothing to do with the variables ${\cal \varphi}$ parametrizing our space ${\cal M}_{2N}$.

Going back to the Hilbert bundle ${\cal V}_{\cal \varphi}$ a section is locally given by a function $\psi$
\begin{eqnarray}
&&\psi:{\cal M}_{2N} \rightarrow {\cal V}\nonumber\\
&&{\cal \varphi} \rightarrow  |\psi;{\cal \varphi}\rangle\in{\cal V}_{\cal \varphi}. \label{6-12}
\end{eqnarray}
Here the notation $|\ldots;{\cal \varphi}\rangle$ indicates that this vector lives in the local Hilbert space ${\cal V}$
(fiber)
associated to the point ${\cal \varphi}$ of the base manifold. At the level of matrix elements the function (\ref{6-12}) is
defined by the components
\begin{equation}
\psi^x({\cal \varphi})=\langle x|\psi;{\cal \varphi}\rangle.
\end{equation}
By replacing ${\cal V}$ by its Hilbert dual ${\cal V}^*$ we arrive at the dual of the Hilbert bundle
\begin{equation}
{\cal X}_x({\cal \varphi})=\langle {\cal X;\varphi}|x\rangle, \qquad \langle {\cal X;\varphi}|\in {\cal V}^*_{\cal
\varphi}.
\end{equation}
In our formalism it is natural also to consider ``multispinor" fields
\begin{equation}
{\cal \varphi}\rightarrow {\cal X}^{x_1\ldots x_q}_{y_1\ldots y_p}({\cal \varphi}) \label{6-13-a}
\end{equation}
which assume values in the tensor product 
\begin{equation}
\underbrace{{\cal V}_{\cal \varphi}^*\otimes {\cal V}_{\cal \varphi}^*\otimes\ldots\otimes {\cal V}_{\cal \varphi}^*}_
{p \;\text{factors}}
\otimes\underbrace{{\cal V}_{\cal \varphi}\otimes{\cal V}_{\cal \varphi}\otimes\ldots\otimes{\cal V}_{\cal \varphi}}_{q\; 
\text{factors}}. \label{6-13-b}
\end{equation}
The symplectic spinors and multispinors have been first studied in great details in Ref. \cite{konstant}. Restricting
ourselves to a spinor, its evolution equation under the Hamiltonian vector field\break
$h^a=\omega^{ab}\partial_bH$ is:
\begin{eqnarray}
\displaystyle \partial_t{\cal X}_x({\cal \varphi},t)&=&-{\cal L}_h{\cal X}_x({\cal \varphi},t)=\nonumber\\
&=&-\int dy \biggl[\delta(x-y)h^a\partial_a+\frac{i}{2}K_{ab}({\cal \varphi})(\Sigma_{\text{meta}}^{ab})^y_{\;x}\biggr]
\cdot {\cal X}_y({\cal
\varphi}, t) \label{6-13}
\end{eqnarray}
where the $K_{ab}(\Sigma^{ab}_{\text{meta}})^y_{\;x}$ is given by (\ref{6-11-b}). 
For notational simplicity we will replace the continuous
index ``$x$" with a discrete one indicated with Greek letters $\alpha$ so that Eq. (\ref{6-13}) is replaced by 
\begin{eqnarray}
\displaystyle \partial_t{\cal X}_{\alpha}({\cal \varphi}, t)&=&-{\cal L}_h{\cal X}_{\alpha}({\cal
\varphi},t)=\nonumber\\
&=& -\biggl[\delta_{\alpha}^{\beta}h^a\partial_a+\frac{i}{2}K_{ab}({\cal \varphi})(\Sigma^{ab})_{\;\alpha}^{\beta}\biggr]
{\cal X}_{\beta}({\cal \varphi},t) \label{6-14}
\end{eqnarray}
where we have only made sure that the group (or manifold) indices are in Latin letters $(a,b)$ and the representation
index are in Greek letters $(\alpha,\beta)$. Note also that we have not put the label ``meta" on the $\Sigma^{ab}$
matrix just because (\ref{6-14}) is the form this equation would have in any representation. 

\section{Metaplectic Hamiltonian and Scalar Product}

Up to now we have used the abstract differential geometric formalism one can find in the literature \cite{marsden} 
\cite{eguchi} \cite{konstant} \cite{dewitt2}, but now we would like to put it in the kind of language we use in Ref.
\cite{Gozzi}. There basically we enlarge the space ${\cal M}_{2N}$ by adding to it $6N$ extra variables
$(\lambda_a,c^a,\bar{c}_a)$ and by making a {\it correspondence} between antisymmetric tensor fields on ${\cal M}_{2N}$
and functions on this enlarged $8N$-dimensional space. 
We call \cite{Gozzi} this {\it correspondence} hat map: ``$\wedge$". We also
realize \cite{Gozzi} that there is an extended Poisson structure on this extended space which turns all the operations of the
standard Cartan calculus (like exterior derivative, interior contraction, etc.) into normal Poisson brackets operations. The
same evolution via a Lie derivative can be turned \cite{Gozzi} into the action of an extended Hamiltonian via the new Poisson
brackets. We would like to build the same formalism in the metaplectic case. The procedure is straightforward \cite{meta}.
Let us extend ${\cal M}_{2N}$ to an ${\cal M}_{\text{ext}}$ space with coordinates $({\cal
\varphi}^a,\lambda_a,\eta^{\alpha},\bar{\eta}_{\alpha})$ where the $\lambda_a$ are the same kind of variables we used in
Ref. \cite{Gozzi} and in the first part of this paper while $\eta^{\alpha},\bar{\eta}_{\alpha}$ are Grassmannian variables
and they are as many as the index $\alpha$. Note that in Ref. \cite{Gozzi} the Grassmannian variables $c^a,\bar{c}_a$ had
the Latin index ``$a$" and so they were as many as the variables ${\cal \varphi}^a$ (or $\lambda_a$). This was because we
were in the vector (or form) representation which has the same dimension as the manifold
${\cal M}_{2N}$. Here instead $\alpha$ can have
the dimension of any representation we are using. So the dimension of ${\cal M}_{\text{ext}}$ is not $8N$ but $4N+4M$ where
$M$ is the dimension of the representation. 

Next let us endow ${\cal M}_{\text{ext}}$ with the following extended Poisson structure (epb):
\begin{eqnarray}
&&\{{\cal\varphi}^a,\lambda_b\}_{epb}=\delta_b^a,\qquad \{{\cal \varphi}^a,{\cal
\varphi}^b\}_{epb}=\{\lambda_a,\lambda_b\}_{epb}=0\nonumber\\
&&\{\bar{\eta}_{\beta},\eta^{\alpha}\}_{epb}=-i\delta^{\alpha}_{\beta},\qquad
\{\eta^{\alpha},\eta^{\beta}\}_{epb}=\{\bar{\eta}_{\alpha},\bar{\eta}_{\beta}\}_{epb}=0 \label{7-1}
\end{eqnarray}
and with the following Hamiltonian
\begin{equation}
\displaystyle \widetilde{\cal H}_{\scriptscriptstyle \text{MFA}}=h^a({\cal \varphi})
\lambda_a+\frac{1}{2}\bar{\eta}_{\alpha}
K_{ab}({\cal \varphi})(\Sigma^{ab}_{\text{meta}})^{\alpha}_{\;\beta}\eta^{\beta}. \label{7-2}
\end{equation}
As we said in the Introduction, the acronym {\it MFA} means ``\underline{M}etaplectic \underline{F}unctional 
\underline{A}pproach" and we have used it 
also for this Hamiltonian because it will be the one appearing in the functional approach which we will present 
later on.
As last ingredient let us build the hat ``$\wedge$" map we mentioned above \cite{Gozzi} between multispinor fields
(\ref{6-13-a}) of the abstract formalism and objects belonging to ${\cal M}_{\text{ext}}$.
The ``$\wedge$-map" is
\begin{equation}
\displaystyle {\cal X}({\cal \varphi})\;\rightarrow\;
\widehat{\cal X}\equiv \frac{1}{p!q!}{\cal X}^{\beta_1\ldots\beta_q}_{\alpha_1\ldots\alpha_p}({\cal \varphi})
\bar{\eta}_{\beta_1}\ldots\bar{\eta}_{\beta_q}\eta^{\alpha_1}\ldots\eta^{\alpha_p}. \label{7-3}
\end{equation}
It is then a straightforward but very long calculation to show that the Lie derivative of
${\cal X}$ is realized as the extended Poisson bracket with
$\widetilde{\cal H}_{\scriptscriptstyle \text{MFA}}$
\begin{equation}
({\cal L}_h{\cal X})\;\hat{\longrightarrow}\;-\{\widetilde{\cal H}_{\scriptscriptstyle \text{MFA}},
\widehat{\cal X}\}_{epb}.
\label{7-4}
\end{equation}
For example it is easy to show that from the following equation: 
\begin{equation}
\partial_t\widehat{\cal X}=\{\widetilde{\cal H}_{\scriptscriptstyle \text{MFA}},\widehat{\cal X}\}_{epb} \label{7-5}
\end{equation}
where $\widehat{\cal X}$ is the $\wedge$-correspondent of a spinor field,
one gets, by stripping it of the $\eta^{\alpha}$ field, the standard equation for the spinor field (\ref{6-14})
\begin{equation}
\partial_t{\cal X}_{\alpha}=-{\cal L}_h{\cal X}_{\alpha}.
\end{equation}
Via $\widetilde{\cal H}_{\scriptscriptstyle MFA}$ and the extended Poisson brackets it is easy to obtain
the evolution of all variables $({\cal \varphi}^a,\lambda_a,\eta^{\alpha},
\bar{\eta}_{\alpha})$ of our manifold ${\cal M}_{\text{ext}}$. For ${\cal \varphi}^a$ the equation of motion one gets is the
same as the standard one \cite{Gozzi} of classical mechanics
\begin{equation}
\dot{\cal \varphi}^a=h^a({\cal \varphi}(t)) \label{7-6}
\end{equation}
while for the Grassmannian variables they are 
\begin{equation}
\left\{
	\begin{array}{l}
	\displaystyle 
	\dot{\eta}^{\alpha}=-\frac{i}{2}K_{ab}(\Sigma^{ab}_{\text{meta}})^{\alpha}_{\;\beta}\eta^{\beta} \smallskip\\
	\displaystyle \dot{\bar{\eta}}_{\alpha}=\frac{i}{2}K_{ab}\bar{\eta}_{\beta}
	(\Sigma^{ab}_{\text{meta}})^{\beta}_{\;\alpha}
	\label{7-7}.
	\end{array}
	\right.
\end{equation}
Let us notice that the last two equations are quite different from the one of the Jacobi field $\delta{\cal \varphi}^a$
\begin{equation}
\displaystyle \frac{d}{dt}(\delta{\cal \varphi}^a)=\partial_lh^a({\cal \varphi})(\delta{\cal \varphi}^l).
\label{7-8}
\end{equation}
So we cannot identify the $\eta^{\alpha}$ with the Jacobi fields of classical mechanics. What are they? It is easy to
show that they are a sort of ``square root" of the Jacobi fields \cite{meta} in the sense that composite objects
${\mathscr P}^a(t)$ defined as 
\begin{equation}
{\mathscr P}^a(t)\equiv\bar{\eta}_{\alpha}(\gamma^a)^{\alpha}_{\;\beta}\bar{\eta}^{\beta} \label{7-9}
\end{equation}
have the same equations of motion as the Jacobi fields. The details of this derivation are given in Appendix G.

The extended Poisson brackets formalism presented in formulae
(\ref{7-1})-(\ref{7-5}) can be given a classical path integral version
as explained in details in Ref. \cite{meta}. The associated generating functional is
\begin{equation}
\displaystyle Z_{\scriptscriptstyle \text{MFA}}=\int {\mathscr D}{\cal \varphi}{\mathscr D}\lambda{\mathscr D}
\eta{\mathscr D}\bar{\eta}
\;\text{exp}\,i\int dt\bigl[\lambda_a\dot{\varphi}^a+i\bar{\eta}_{\alpha}\dot{\eta}^{\alpha}-
\widetilde{\cal H}_{\scriptscriptstyle \text{MFA}}\bigr].
\label{7-10}
\end{equation}
As we did for the {\it CPI} case \cite{Gozzi} it is easy to derive the ``operatorial" version of this {\it MFA} path
integral. From the kinetic term in (\ref{7-10}) one gets the following commutators
\begin{equation}
\left\{
	\begin{array}{l}
	\displaystyle 
	[\widehat{\cal\varphi}^a,\widehat{\lambda}_b]=i\delta_b^a \smallskip\\
	\displaystyle [\widehat{\bar{\eta}}_{\beta},\widehat{\eta}^{\alpha}]=\delta^{\alpha}_{\beta}
	\label{7-11}
	\end{array}
	\right.
\end{equation}
where with $[(\cdot),(\cdot)]$ we indicated $Z_2$-graded commutators. All other commutators not indicated in (\ref{7-11})
are zero. 
In a ``Schr\"odinger-type" representation where $\widehat{\cal \varphi}^a$ and $\widehat{\eta}^{\alpha}$ are 
{\it multiplicative} operators, the associated momenta operators $\widehat{\lambda}_a$, 
$\widehat{\bar{\eta}}_{\alpha}$ have to be 
realized as {\it derivative} operators in order to satisfy the algebra (\ref{7-11})
\begin{eqnarray}
\displaystyle &&\widehat{\lambda}_a=-i\frac{\partial}{\partial{{\cal \varphi}^a}}\nonumber\\
&&\widehat{\bar{\eta}}_{\alpha}=\frac{\partial}{\partial\eta^{\alpha}}. \label{7-12}
\end{eqnarray}
The representation space on which $\widehat{\cal \varphi}^a,\widehat{\eta}^{\alpha}$ are represented as multiplicative operators is
given by the set of functions 
\begin{equation}
\displaystyle {\cal X}({\cal \varphi},\eta)\equiv \sum_p\frac{1}{p!}{\cal X}_{\alpha_1\alpha_2\ldots\alpha_p}
({\cal \varphi}) \eta^{\alpha_1}\eta^{\alpha_2}\ldots \eta^{\alpha_p} \label{7-13}
\end{equation}
and the metaplectic Hamiltonian (\ref{7-2}) is turned into the operator
\begin{equation}
\displaystyle \widehat{\cal H}_{\scriptscriptstyle \text{MFA}}=\widetilde{\cal H}_{\scriptscriptstyle \text{MFA}}
\biggl(\widehat{\cal \varphi},\widehat{\lambda}=-i\frac{\partial}{\partial{\cal \varphi}},
\widehat{\eta},\widehat{\bar{\eta}}=\frac{\partial}{\partial\eta}\biggr). \label{7-14}
\end{equation}
As explained in Ref. \cite{meta} for the metaplectic Hamiltonian there are some ordering ambiguities. From now on, like in
Ref. \cite{meta}, we will choose an ``anti-normal" ordering, i.e. the operator $\displaystyle \widehat{\bar{\eta}}
=\frac{\partial}{\partial \eta}$ will act always to the right of $\widehat{\eta}$. 

The next step is to endow the space of functions (\ref{7-13}) with a scalar product and check if the $\widehat{\cal
H}_{\scriptscriptstyle \text{MFA}}$ is Hermitian under it. 
The scalar product we will choose is the analog of the SvH one introduced in Ref. \cite{one} for the {\it CPI} case. 
The analog of the hermiticity
conditions for the SvH case were
\begin{equation}
\left\{
	\begin{array}{l} 
	\eta^{\alpha\dagger}=\bar{\eta}_{\alpha} \smallskip \\
	\bar{\eta}_{\alpha}^{\dagger}=\eta^{\alpha} \smallskip\\
	{\cal \varphi}^{a\dagger}={\cal \varphi}^a \smallskip\\
	\lambda_a^{\dagger}=\lambda_a.
	\label{7-15}
	\end{array}
	\right.
\end{equation}
Along the same lines we have developed in Ref. \cite{one}, it is easy to show that the scalar product induced by the hermiticity
conditions (\ref{7-15}) among the states (\ref{7-13}) is
\begin{equation}
\displaystyle \langle \tau|{\cal X}\rangle=\sum_pK(p)\tau^{*\alpha_1\ldots \alpha_p}({\cal \varphi})
{\cal X}_{\alpha_1\ldots\alpha_p}
({\cal \varphi}). \label{7-16}
\end{equation}
where $K(p)$ is a positive combinatorial factor.
One immediately notices that this is a {\it positive definite} scalar product. The SvH scalar product is positive
definite also in the {\it CPI} case \cite{one}.
Let us now check whether the Hamiltonian $\widehat{\cal H}_{\scriptscriptstyle \text{MFA}}$ is Hermitian. In the {\it CPI} 
case \cite{one}
it is not. Let us first remember that the bosonic part of $\widehat{\cal H}_{\scriptscriptstyle \text{MFA}}$ (\ref{7-2}), 
which is the same as
in the {\it CPI} case, is Hermitian \cite{one}. So we have to check out only the Fermionic (or Grassmannian) part which is
\begin{equation}
\displaystyle \widehat{\cal
H}_{\scriptscriptstyle \text{MFA}}^{\text{ferm}}=\frac{1}{2}\partial_a\partial_bH\bar{\eta}_x
(\Sigma^{ab}_{\text{meta}})^x_{\; y}\eta^y.
\label{7-17}
\end{equation}
We have indicated the indices with $x,y$ because, in the metaplectic case, 
they are a continuous set of indices as explained
previously. They label in fact the infinite states of the Hilbert space ${\cal V}$.
Second, let us indicate the elements $(\Sigma^{ab}_{\text{meta}})^x_{\;y}$ as 
$\langle x|\Sigma^{ab}_{\text{meta}}|y\rangle$ and let us remember that
they had to be chosen Hermitian in the metaplectic and in any unitary representation
\begin{equation}
(\Sigma^{ab}_{\text{meta}})^{\dagger}=\Sigma^{ab}_{\text{meta}}. \label{7-18}
\end{equation}
This hermiticity of course refers to the indices $(x,y)$ and not to $(a,b)$. As a consequence of (\ref{7-18}) we have
\begin{equation}
\langle x|\Sigma^{ab}_{\text{meta}}|y\rangle^*=\langle y|\Sigma^{ab\;\dagger}_{\text{meta}}|x\rangle
=\langle y|\Sigma_{\text{meta}}^{ab}|x\rangle
\end{equation}
which in normal matrix language means
\begin{equation}
(\Sigma^{ab}_{\text{meta}})^{x*}_{\;y}=(\Sigma^{ab}_{\text{meta}})^y_{\;x}. \label{7-19}
\end{equation}
Let us now check the hermiticity of $\widehat{\cal H}_{\scriptscriptstyle \text{MFA}}^{\text{ferm}}$ written in (\ref{7-17}):
\begin{eqnarray}
(\widehat{\cal H}^{\text{ferm}}_{\scriptscriptstyle \text{MFA}})^{\dagger}&=&\biggl(\frac{1}{2}
(\partial_a\partial_bH)\bar{\eta}_x(\Sigma^{ab}_{\text{meta}})^x_{\;y}\eta^y\biggr)^{\dagger}=\nonumber\\
&=&\frac{1}{2}(\partial_a\partial_bH)\eta^{y\dagger}(\Sigma^{ab}_{\text{meta}})^{x*}_{\;y}\bar{\eta}_x^{\dagger}=
\nonumber\\
&=&\frac{1}{2}(\partial_a\partial_bH)\bar{\eta}_y(\Sigma^{ab}_{\text{meta}})^y_{\;x}\eta^x=\nonumber\\
&=&\frac{1}{2}(\partial_a\partial_bH)\bar{\eta}_x(\Sigma^{ab}_{\text{meta}})^x_{\;y}\eta^y=\widehat{\cal
H}_{\scriptscriptstyle \text{MFA}}^{\text{ferm}}.
\end{eqnarray}
In the third step above we made use of the SvH hermiticity conditions (\ref{7-15}) for the $\eta^x$, $\bar{\eta}_x$
and of the relation (\ref{7-19}). So this proves that the full $\widetilde{\cal H}_{\scriptscriptstyle \text{MFA}}$ 
is Hermitian under the
SvH scalar product. This does not happen for the $\widetilde{\cal H}$ 
of the {\it CPI} \cite{Gozzi}\cite{one}. 
Let us understand why. It was shown in Appendix F that also the usual $\widetilde{\cal
H}_{\scriptscriptstyle CPI}$ can be given a form similar to 
$\widetilde{\cal H}_{\scriptscriptstyle \text{MFA}}$:
\begin{equation}
\widetilde{\cal H}_{\scriptscriptstyle CPI}=
h^a\lambda_a+\frac{1}{2}\bar{c}_eK_{ab}({\cal \varphi})(\Sigma^{ab}_{\text{vec}})^e_{\;f}
c^f
\end{equation}
where $(\Sigma^{ab}_{\text{vec}})^e_f$ is the $\Sigma$ associated to the transformations of vectors 
under $\text{Sp}(2N)$ (see (\ref{e-3})) and is given by:
\begin{equation}
(\Sigma^{ab}_{\text{vec}})^e_{\; f}=-i(\delta^a_f\omega^{be}+\delta^b_f\omega^{ae}).
\end{equation}
It is easy to check that this $\Sigma$ does not satisfy the analog of the relation (\ref{7-17}) that means
\begin{equation}
(\Sigma^{ab}_{vec})^{e*}_{\;f}\neq (\Sigma^{ab}_{\text{vec}})^f_{\; e}.
\end{equation}
This explains why $\widetilde{\cal H}_{\scriptscriptstyle CPI}$ is not Hermitian
in the SvH scalar product. 

There may be other scalar products in the metaplectic case which are both positive definite and under which $\widetilde{\cal
H}_{\scriptscriptstyle \text{MFA}}$ is Hermitian but for the moment we have not initiated any search for them. This search anyhow may be needed
in the future as explained in the Conclusions. 

\section{Conclusions and Outlook}

In this paper we have analyzed two new operatorial extensions of the Koopman-von
Neumann ({\it KvN}) 
approach which, differently 
from the standard {\it CPI} case studied in Ref. \cite{one}, present both a
Hermitian Hamiltonian and a positive definite
scalar product. Leaving for a moment aside the metaplectic case ({\it MFA}) let
us concentrate 
on the bosonic one ({\it BFA}). The reader may prefer this one over the {\it
CPI} case but there are several
drawbacks we want to point out. First of all in the {\it BFA} approach the
higher tensors and forms 
had to be built by hand introducing from the outside the operation $\otimes$ of
tensor product 
(\ref{tensorpr}), while in the {\it CPI} case, because of the Grassmannian
nature of the 
variables $c$, the higher tensors and forms were generated automatically as
functions 
on the extended phase space which is the sole ingredient entering the associated
path integral.
Moreover at the operatorial level in the {\it BFA} case we had to build several
copies (\ref{5-3}) of the basic 
Hilbert space in order to get the higher tensors and forms. As a consequence the
associated path integral 
(\ref{5-12-a}) is quite awkward and it does not have a simple interpretation in
terms of Dirac deltas on the classical
trajectories. More serious than this drawback is another one that we fear may
affect the {\it BFA}.
It concerns the following problem. We have seen in Ref. \cite{one} that the
non-hermiticity of $\widetilde{\cal H}
_{\scriptscriptstyle CPI}$
or the non-positive definiteness of the scalar product were crucial ingredients
in order to describe 
chaotic systems. In fact such ingredients can imply the presence of complex
eigenvalues 
for $\widetilde{\cal H}_{\scriptscriptstyle CPI}$ and this has as a consequence
the exponential 
increase of the Jacobi fields. 
Nothing like that can happen with the $\widetilde{\cal H}_{\scriptscriptstyle
BFA}$
which is Hermitian and with a positive definite scalar product. Does it mean
that 
$\widetilde{\cal H}_{\scriptscriptstyle BFA}$ 
cannot describe all systems? We feel it will but most probably we will have to
further enlarge the Hilbert
space of the {\it BFA}. People have gone in this direction already with other
Hermitian operators.
For example the authors of Ref. \cite{benatti}, in order to get the chaotic
behavior out of the analog 
of the Hermitian {\it KvN} operator for zero-forms, enlarged the Hilbert space
to a rigged Hilbert space 
where the operator was no longer Hermitian. This may be the road to pursue also
in the {\it BFA} case. Before 
doing that anyhow one should really check whether this further extension to a
rigged Hilbert space 
is needed or if some mathematical subtleties of the {\it BFA} allow us to
describe also chaotic systems 
without any further extension. We have not embarked on this study but we have,
in this paper, prepared
the mathematical ground to do that by analyzing in all details the geometry
underlying the {\it BFA}.
That the {\it CPI} instead could describe chaotic systems was not only indicated
in Ref. \cite{one}
by the presence of complex eigenvalues of $\widetilde{\cal
H}_{\scriptscriptstyle CPI}$ 
but it was shown explicitly in Ref. \cite{fractals}
where an explicit expression of the Lyapunov exponents in terms of the {\it CPI}
generating functional 
$Z_{\scriptscriptstyle CPI}[J]$ was written down. 

Let us now turn to the metaplectic case. Why did we study it here? We did first
of all to present 
another example of an extension of the {\it KvN} zero-form formalism which has
both a Hermitian 
Hamiltonian and a positive definite scalar product. These mathematical features
were not studied 
in the first presentation \cite{meta} of the {\it MFA}. Of course for this
model, differently 
than the {\it CPI} and the {\it BFA}, we do not have in mind applications to
chaotic systems
or similar things; what we have in mind is the light it may throw on the issue
of quantization.
It was used in that respect in Ref. \cite{reutermeta}. There quantization was
achieved 
by first postulating a {\it MFA} dynamics for the extended {\it KvN} dynamics
and next introducing
a flat connection on the Hilbert bundle defined in {\bf Sec. IV}. The main thing
we want to understand of that project is why we need to start at the classical
level 
from the {\it MFA} dynamics. An answer to this question that we are currently
exploring is the following.
Maybe the {\it CPI} should be considered the right classical dynamics not for
the classical
wave functions but for the {\it probability densities}
\begin{equation}
\rho(\varphi,c)=\rho_0(\varphi)+\rho_a(\varphi)c^a+\rho_{ab}(\varphi)c^ac^b+\ldots
\label{A}
\end{equation}
which are only integrable (i.e. belonging to $L^1$) and not square integrable
functions.
Then to get the ``classical wave functions" we should do a sort of ``square
root" 
of the $\rho$ in (\ref{A}). May it be that these ``square roots" are the {\it
MFA} wave functions?
\begin{equation}
\displaystyle \label{A1} \psi(\varphi,\eta)=\sum_p\frac{1}{p!}\psi_{a_1\ldots
\alpha_p}(\varphi)\eta^{\alpha_1}
\eta^{\alpha_2}\ldots \eta^{\alpha_p}
\end{equation} 
If so this would explain why we need the {\it MFA} evolution at the classical
level.
The reason we have this suspect is because the $\eta$ are something like the 
``square roots" (\ref{7-9}) of the $c$:
\begin{equation}
c^a=\bar{\eta}_{\alpha}(\gamma^a)^{\alpha}_{\;\beta}\bar{\eta}^{\beta}.
\end{equation}
What we actually need in the {\it MFA} is a new scalar product such that
\begin{equation}
_{\cal \varphi}\langle \psi|\eta\rangle\langle\eta|\psi\rangle_{\cal \varphi}=
\rho({\cal \varphi},\bar{\eta}\gamma\eta)=\rho({\cal \varphi},c).
\label{8-3}
\end{equation}
That means that we would like that the $\eta,\bar{\eta}$ on the LHS of
(\ref{8-3}) 
get combined by this scalar product into those combinations
$\bar{\eta}\gamma^a\eta$ which
are basically the $c^a$. We want that they combine in this way because the
classical 
probability densities in (\ref{A}) contain the forms $c$ and not $\eta$ or
$\bar{\eta}$.
The scalar product (\ref{8-3}) is not the SvH one that we explored in {\bf Sec.
V} for the {\it MFA}.
In fact the SvH scalar product of the {\it MFA} does not pull in the $\gamma$
matrices which instead 
are necessary in (\ref{8-3}) to get the combination $\bar{\eta}\gamma\eta$
inside the $\rho$.
So far we have not succeeded in building the scalar product (\ref{8-3}) but in
order to get some
practice we have asked ourselves how, from the various components of the $\psi$
appearing in 
(\ref{A1}), we can build objects which at least have the same indices and
transformation
properties as the various components of $\rho$ appearing in (\ref{A}). One
solution we found (see Appendix H) is the following one:
\begin{equation}
\rho_{\underbrace{\scriptstyle ab\ldots d}_{p}}({\cal
\varphi})={\text{Tr}}\bigl[|\psi^{\scriptscriptstyle (p)}
\rangle\langle\psi^{\scriptscriptstyle (p)}|\gamma_{[a}\otimes\gamma_b
\otimes\ldots\gamma_{d]}\bigr] \label{8-4}
\end{equation}
where with $|\psi^{\scriptscriptstyle (p)}\rangle$ we indicate the components of
the states (\ref{7-13}) with $p$ indices
while with $\otimes$
we indicate the tensor product among the Hilbert spaces like in (\ref{6-13-b}). 
The {\it first} thing to notice in (\ref{8-4}) is
that if we transform $|\psi^{\scriptscriptstyle (p)}\rangle$ 
according to the metaplectic transformations then the resulting $\rho_{ab\ldots
d}$
turns out to transform according to the symplectic one. 

{\it Second}, let us notice that
$|\psi^{\scriptscriptstyle (p)}\rangle$ has, in the metaplectic case, components
$\psi_{\alpha_1\ldots\alpha_p}$ whose
number of indices can
run from zero to $\infty$, while $\rho$ can have only at most $N$ indices. This
means we have much more information
stored in the $|\psi\rangle$ that what is needed to build the $\rho$. What does
this mean? 

{\it Third}, let us remember that
while (\ref{8-4}) produces a $\rho$ out of a $\psi$, it is not clear whether the
inverse procedure 
is true and unique. That means whether,
given a $\rho$ with all its components, it is possible to find a $|\psi\rangle$
such that (\ref{8-4}) or (\ref{8-3})
is satisfied. 

This is the project we are currently working and this explains why it is crucial
for this project to analyze 
the various scalar products associated to the {\it MFA} dynamics.
\begin{acknowledgments}
\noindent
We wish to thank M. Reuter for many useful discussions. 
The work of E.D. has been supported by an INFN Postdoctoral
Fellowship at MIT. He wishes to thank the members of the Center for Theoretical Physics at MIT for hospitality. 
The work of E.G.
and D.M. has been supported in part by funds from INFN, MIUR and the University of Trieste.
The research contained in this paper and in \cite{one} has been re-triggered by the appearance of Ref. 
\cite{Marnelius}. We wish to thank R. Marnelius for some useful e-mail messages.
\end{acknowledgments}

\appendix
\makeatletter
\@addtoreset{equation}{section}
\makeatother
\renewcommand{\theequation}{\thesection\arabic{equation}}


\newpage

\section{}
\noindent

In this Appendix we give a proof of (\ref{2-3}). Even if quite formal, we hope it will convince the reader of the correctness
of Eq. (\ref{2-3}). The determinants in (\ref{2-3}) are functional determinants that means:
\begin{eqnarray}
\displaystyle \text{det}\biggl[\delta_l^a\partial_t-\omega^{ab}\frac{\partial^2H}{\partial\varphi^b\partial\varphi^l}\biggr]
&\equiv& \text{det}\biggl[\delta_l^a\partial_t\delta(t-t^{\prime})-\delta(t-t^{\prime})\,\omega^{ab}\frac{\partial^2H}
{\partial\varphi^b\partial\varphi^l}\biggr]=\nonumber\\
&=&\{\text{det}\,\partial_t\}\biggl\{\text{det}\biggl[\delta_l^a\delta(t-t^{\prime})-\theta(t-t^{\prime})\,
\omega^{ab}\frac{\partial^2H}
{\partial\varphi^b\partial\varphi^l}\biggr]\biggr\}. \label{a-1}
\end{eqnarray}
The proof of (\ref{2-3}) is equivalent to the statement that the determinant of the product of the two matrices entering
respectively the RHS and LHS of (\ref{2-3}) is one. To prove that, let us use the form of the matrix written in the second
line of (\ref{a-1}) and let us drop the factor $(\text{det}\,\partial_t)$ which is constant (independent of $\varphi$). 
What we get, as determinant of the product of the two matrices, is:
\begin{eqnarray}
\displaystyle && \text{det}\biggl\{\int dt^{\prime}\biggl[\delta_b^a\delta(t-t^{\prime})-\theta(t-t^{\prime})\,
\omega^{al}\frac{\partial^2H}{\partial\varphi^l\partial\varphi^b}\biggr]\biggl[\delta_c^b\delta(t^{\prime}-t^{\prime
\prime})
+\theta(t^{\prime}-t^{\prime\prime})\omega^{bk}\frac{\partial^2H}{\partial\varphi^k\partial\varphi^c}\biggr]
\biggr\}=\nonumber\\
&&=\text{det}\biggl\{\delta_c^a\delta(t-t^{\prime\prime})-\int dt^{\prime}\theta(t-t^{\prime})\theta(t^{\prime}-
t^{\prime\prime})\,
\omega^{al}\frac{\partial^2H}{\partial\varphi^l\partial\varphi^b}\cdot\omega^{bk}\frac{\partial^2H}
{\partial\varphi^k\partial\varphi^c}\biggr\} \label{a-2}\\
&&\approx \text{exp}\biggl(-\int dt^{\prime}\,
\theta(t-t^{\prime})\theta(t^{\prime}-t)\,\omega^{al}\frac{\partial^2H}{\partial\varphi^l\partial\varphi^b}\omega^{bk}
\frac{\partial^2H}{\partial\varphi^k\partial\varphi^a}\biggr)=1. \label{a-3}
\end{eqnarray}
In (\ref{a-3}) we have used the ``exp-tr" form for the determinant and the fact that the product of the two $\theta(\cdot)$
present in (\ref{a-3}) gives zero. In (\ref{a-1}) and (\ref{a-2}) we have used the $\theta(t-t^{\prime})$ as 
``inverse" (or Green function) of the operator $\partial_t$. The function 
$\theta(t-t^{\prime})$ is actually the {\it retarded}
or causal Green function. If we had used other Green functions, like for example $\epsilon(t^{\prime}-t)$, we would
not have obtained the result (\ref{a-3}). The reason to use ``causal" boundary conditions is because, after all, the
determinants above are related to the standard Hamilton equations of motion and these are usually solved by giving a value of
$\varphi$ at the initial time $t=0$ and by determining the evolution at later times using a causal propagator.

If the reader is not convinced by our {\it formal} proof presented in (\ref{a-3}), we will now present a new one. It is
actually well known \cite{nakazato} that all the functional determinants of the form
\begin{equation}
\text{det}[\partial_t\delta(t-t^{\prime})-\delta(t-t^{\prime})G^{\prime}(\varphi)] \label{a-4}
\end{equation}
depend on the {\it boundary conditions} under which we solve the associated differential equation
\begin{equation}
[\partial_t-G^{\prime}(\varphi)]c_n(t)=\sigma_nc_n(t) \label{a-5}
\end{equation}
whose eigenvalues $\sigma_n$ are needed to calculate the determinant in some {\it regularized} form:
\begin{equation}
\text{det}[(\;\;\;)]=\Biggl\{\prod_{n=-\infty}^{n=\infty}\sigma_n\Biggr\}_{\text{regul}}. \label{a-6}
\end{equation}
Solving Eq. (\ref{a-5}) with {\it causal} boundary conditions, one obtains \cite{nakazato} from (\ref{a-6}) 
\begin{equation}
\displaystyle \text{det}[\partial_t\delta(t-t^{\prime})-\delta(t-t^{\prime})G^{\prime}(\varphi)]_{\text{causal}}=
\text{exp}\biggl(-\frac{1}{2}
\int dt^{\prime} G^{\prime}\bigl(\varphi(t^{\prime})\bigr)\biggr). \label{a-7}
\end{equation}
By changing in (\ref{a-7}) the sign of $G^{\prime}(\varphi)$, one gets
\begin{equation}
\displaystyle \text{det}[\partial_t\delta(t-t^{\prime})+\delta(t-t^{\prime})G^{\prime}(\varphi)]_{\text{causal}}
=\text{exp} \biggl(+\frac{1}{2}\int dt^{\prime}G^{\prime}\bigl(\varphi(t^{\prime})\bigr)\biggr). \label{a-8}
\end{equation}
By comparing the RHS of (\ref{a-7}) and (\ref{a-8}) one sees that the two determinants on the LHS are one the inverse of
the other. This proves relation (\ref{a-3}) provided we specify that the functional determinant be evaluated with
{\it causal} boundary conditions. $G^{\prime}(\varphi)$ is in our case $\displaystyle \omega^{ab}\frac{\partial^2H}
{\partial\varphi^b\partial\varphi^l}$, which means is a matrix but the same formulas like (\ref{a-7}) and (\ref{a-8}) hold
also for matrices.

The reader may wonder what happens if one uses, for example, periodic boundary conditions and a time-symmetric Green
function. This has been analyzed in full details in Ref. \cite{reuter} and the result is that, if that determinant is
inserted in (\ref{2-2}), the associated generating functional gives {\it non-zero} expectation values only to those
observables which are independent of deformations of $H$ and of its symplectic form $\omega_{ab}$. This means a path
integral which does not feel anymore the form of $H$. This is
something similar to a topological field theory but that is not what we want here. 


\newpage

\section{}

\noindent In this Appendix we will prove the relation (\ref{4-10}). Using the basic commutators (\ref{3-1}) we get

\begin{eqnarray}
\displaystyle && [i\widehat{\cal H}_{H_1},i\widehat{\cal H}_{H_2}]=-[\lambda_a\omega^{ab}\partial_bH_1-\pi^a\omega^{bc}\partial_c
\partial_aH_1\xi_b,\lambda_d\omega^{de}\partial_eH_2-\pi^d\omega^{ef}\partial_f\partial_dH_2\xi_e]=\nonumber\\
&&=-i\lambda_a\omega^{ab}\partial_b\partial_dH_1\omega^{de}\partial_eH_2+i\lambda_d\omega^{de}\partial_a\partial_eH_2
\omega^{ab}\partial_bH_1-i\omega^{ab}\partial_bH_1\pi^d\omega^{ef}\partial_a\partial_f\partial_dH_2\xi_e
\nonumber\\
&&\quad+i\pi^a\omega^{bc}\partial_c\partial_d\partial_aH_1\xi_b\omega^{de}\partial_eH_2-i\pi^a\omega^{bc}\partial_c
\partial_a
H_1\omega^{ef}\partial_f\partial_bH_2\xi_e+
i\pi^d\omega^{ef}\partial_f\partial_dH_2\omega^{bc}\partial_c\partial_eH_1\xi_b=\nonumber\\
&&=-i\lambda_a\omega^{ab}\partial_b\partial_dH_1\omega^{de}\partial_eH_2-i\lambda_a\omega^{ab}\partial_b\partial_eH_2\omega
^{de}\partial_dH_1+i\pi^d\omega^{ef}\partial_f\partial_d\partial_aH_2\partial_{b}H_1\omega^{ba}\xi_e\nonumber\\
&&\quad +i\pi^a\omega^{bc}\partial_c\partial_a\partial_dH_1\omega^{de}\partial_eH_2\xi_b
+i\pi^a\omega^{ef}\partial_a\partial_cH_1\omega^{cb}\partial_b\partial_fH_2\xi_e+
i\pi^d\omega^{bc}\partial_c\partial_eH_1\omega^{ef}\partial_f\partial_dH_2\xi_b=\nonumber\\
&&=-i[\lambda_a\omega^{ab}\partial_b(\partial_dH_1\omega^{de}\partial_eH_2)-\pi^d\omega^{ef}\partial_f\partial_d
(\partial_bH_1\omega^{ba}\partial_aH_2)\xi_e]=-i\widehat{\cal H}_{\{H_1,H_2\}}.
\end{eqnarray}

\newpage

\section{}

\noindent $\bullet$ \underline{\it Derivation of (\ref{4-15})}.

\noindent In this derivation and in the following we will basically use only the commutation relations (\ref{3-1}) and the
fact that $\omega^{ab}$ is antisymmetric in $(a,b)$. 

\begin{eqnarray}
[Q_g^{\scriptscriptstyle (B)},\widehat{\cal H}]&=&[i\pi^a\xi_a,\lambda_b\omega^{be}\partial_eH-
\pi^e\omega^{db}\partial_b\partial_eH\xi_d]=
\nonumber\\
&=&[i\pi^a\xi_a,-\pi^e\omega^{db}\partial_b\partial_eH\xi_d]=\nonumber\\
&=&\pi^e\omega^{db}\partial_b\partial_eH\xi_d-\pi^e\omega^{db}\partial_b\partial_eH\xi_a\delta_d^a=\nonumber\\
&=&0
\end{eqnarray}

\begin{eqnarray}
[N^{\scriptscriptstyle (B)},\widehat{\cal H}]&=& 
[\pi^a\partial_aH,\lambda_b\omega^{be}\partial_eH-\pi^e\omega^{ab}\partial_b\partial_eH\xi_a]
\nonumber\\
&=&i\pi^a\partial_a\partial_bH\omega^{be}\partial_eH+i\pi^e\omega^{ab}
\partial_b\partial_eH\partial_aH=\nonumber\\
&=&i\pi^a\partial_a\partial_bH\omega^{be}\partial_eH+i\pi^a\omega^{eb}\partial_b\partial_aH\partial_eH=\nonumber\\
&=&0
\end{eqnarray}

\begin{eqnarray}
[\overline{N}^{\scriptscriptstyle (B)},\widehat{\cal H}]&=& 
[\xi_d\omega^{de}\partial_eH,\lambda_b\omega^{bc}\partial_cH-\pi^l
\omega^{ab}\partial_b\partial_lH\xi_a]=\nonumber\\
&=&i\xi_d\omega^{de}\partial_b\partial_eH\omega^{bc}\partial_cH-i\omega^{le}\partial_eH\omega^{ab}\partial_b\partial_lH
\xi_a=\nonumber\\
&=&i\xi_d\omega^{de}\partial_b\partial_eH\omega^{bl}\partial_lH-i\omega^{bl}\partial_lH\omega^{de}\partial_e\partial_bH
\xi_d=\nonumber\\
&=&0
\end{eqnarray}

\bigskip

\noindent $\bullet$ \underline{\it Derivation of (\ref{4-17})}.

\noindent 
\begin{eqnarray}
[\overline{Q}^{\scriptscriptstyle (B)},\widehat{\cal H}]&=&[i\xi_a\omega^{ab}
\lambda_b,\lambda_i\omega^{ij}\partial_jH-\pi^l\omega^{st}(\partial_t
\partial_lH)\xi_s]=\nonumber\\
&=&i\xi_a\omega^{ab}[\lambda_b,\lambda_i\omega^{ij}\partial_jH-\pi^{l}\omega^{st}
(\partial_t\partial_lH)\xi_s]+i[\xi_a,-\pi^l\omega^{st}(\partial_t\partial_lH)\xi_s]\omega^{ab}\lambda_b=\nonumber\\
&=&\xi_a\omega^{ab}\lambda_i\omega^{ij}(\partial_b\partial_jH)-\xi_a\omega^{ab}\xi_s\omega^{st}
(\partial_b\partial_t\partial_lH)\pi^l
+\omega^{st}(\partial_t\partial_lH)\xi_s\omega^{lb}\lambda_b=\nonumber\\
&=&-\xi_a\omega^{ab}\xi_s\omega^{st}(\partial_b\partial_t\partial_lH)\pi^l.
\end{eqnarray}

The derivation of (\ref{4-16}) is as straightforward as the previous one and we will leave it to the reader.

\bigskip

\noindent $\bullet$ \underline{\it Derivation of (\ref{4-23})}.

\begin{eqnarray}
[Q_{\scriptscriptstyle H}^{\scriptscriptstyle (B)},\overline{Q}_{\scriptscriptstyle H}^
{\scriptscriptstyle (B)}]&=&[i\pi^a\lambda_a-\pi^a\partial_aH,i\xi_i\omega^{ij}\lambda_j+\xi_d\omega^{de}\partial_eH]=
\nonumber\\
&=&i\lambda_a[\pi^a,i\xi_i\omega^{ij}\lambda_j]+i\pi^a[\lambda_a,\xi_d\omega^{de}\partial_eH]+
i[\pi^a,\xi_d\omega^{de}\partial_eH]\lambda_a\nonumber\\
&&-\pi^a[\partial_aH,i\xi_i\omega^{ij}\lambda_j]
-[\pi^a,i\xi_i\omega^{ij}\lambda_j]\partial_aH-[\pi^a,\xi_d\omega^{de}\partial_eH]\partial_aH=\nonumber\\
&=& 2\lambda_d\omega^{de}\partial_eH+2\pi^a\omega^{de}\partial_e\partial_aH\xi_d=\nonumber\\
&=& 2\widehat{\cal H}_{\scriptscriptstyle BFA} +\{4\pi^a\omega^{dc}\partial_c\partial_aH\xi_d\}.
\end{eqnarray}

\bigskip

\noindent $\bullet$ \underline{\it Derivation of (\ref{4-25})}.

\begin{eqnarray}
\delta_{Q_1^{(B)}}\varphi^a&=&[\epsilon Q-\epsilon \overline{N},\varphi^a]=[\epsilon i\pi^b\lambda_b-\epsilon
\xi_b\omega^{bc}\partial_cH,\varphi^a]=\nonumber\\
&=&\epsilon\pi^a \\
\delta_{Q_1^{(B)}}\xi_a&=&[\epsilon Q-\epsilon \overline{N},\xi_a]=[\epsilon i\pi^b\lambda_b-\epsilon\xi_b\omega^{bc}
\partial_cH,\xi_a]=\nonumber\\
&=&\epsilon\lambda_a \\
\delta_{Q_1^{(B)}}\pi^a &=&[-\epsilon\xi_b\omega^{bc}\partial_cH,\pi^a]=-i\epsilon\omega^{ac}\partial_cH\\
\delta_{Q_1^{(B)}}\lambda_a &=&[-\epsilon\xi_b\omega^{bc}\partial_cH,\lambda_a]=-i\epsilon\xi_b\omega^{bc}\partial_c
\partial_aH.
\end{eqnarray}

\newpage

\section{}

In this Appendix we prove some formulae contained in Sec. V.

\bigskip

\noindent $\bullet$ \underline{\it Derivation of (\ref{5-8})}. 

\begin{eqnarray}
\displaystyle [i\widehat{\cal H},\widehat{F}]&=&
\biggl[i{\lambda}_a\omega^{ab}\partial_bH\otimes{\bf 1}_{\scriptscriptstyle (1)}\otimes{\bf 1}_{\scriptscriptstyle (2)}
-i\omega^{be}\partial_e\partial_aH\otimes(\pi^a_{\scriptscriptstyle (1)}
\xi_b^{\scriptscriptstyle (1)}\otimes{\bf 1}_{\scriptscriptstyle (2)}+{\bf 1}_{\scriptscriptstyle (1)}
\otimes\pi^a_{\scriptscriptstyle (2)}\xi_b^{\scriptscriptstyle (2)}),
\nonumber\\
&& F_{de}\otimes\frac{\pi^d_{\scriptscriptstyle (1)}}{2}\otimes\pi^e_{\scriptscriptstyle (2)}-
F_{de}\otimes\frac{\pi^e_{\scriptscriptstyle (1)}}{2}\otimes\pi_{\scriptscriptstyle (2)}^d\biggr]=\nonumber\\
&=& [i\lambda_a\omega^{ab}\partial_bH,F_{de}]\otimes\frac{\pi^d_{\scriptscriptstyle (1)}}{2}\otimes
\pi^e_{\scriptscriptstyle (2)}-[i\lambda_a\omega^{ab}\partial_bH,F_{de}]\otimes
\frac{\pi^e_{\scriptscriptstyle (1)}}{2}\otimes\pi^d_{\scriptscriptstyle (2)}\nonumber\\
&&-i\omega^{bc}\partial_c
\partial_aHF_{de}\otimes\biggl[\pi^a_{\scriptscriptstyle (1)}\xi_b^{\scriptscriptstyle (1)},
\frac{\pi^d_{\scriptscriptstyle (1)}}{2}\biggr]\otimes\pi^e_{\scriptscriptstyle (2)}
+i\omega^{bc}\partial_c\partial_aHF_{de}\otimes\biggl[\pi^a_{\scriptscriptstyle (1)}\xi_b^{\scriptscriptstyle (1)},
\frac{\pi^e_{\scriptscriptstyle (1)}}{2}\biggr]\otimes
\pi^d_{\scriptscriptstyle (2)}\nonumber\\
&&-i\omega^{bc}\partial_c\partial_aHF_{de}\otimes\frac{\pi^d_{\scriptscriptstyle (1)}}{2}\otimes[\pi^a_{\scriptscriptstyle 
(2)}\xi_b^{\scriptscriptstyle (2)},\pi^e_{\scriptscriptstyle (2)}]
+i\omega^{bc}\partial_c\partial_aHF_{de}\otimes\frac{\pi^e_{\scriptscriptstyle (1)}}{2}\otimes 
[\pi^a_{\scriptscriptstyle (2)}\xi_b^{\scriptscriptstyle (2)},\pi^d_{\scriptscriptstyle (2)}]=\nonumber\\
&=&\omega^{ab}\partial_bH\partial_aF_{de}\otimes\biggl[\frac{1}{2}({\pi}^d_{\scriptscriptstyle (1)}
\otimes{\pi}^e_{\scriptscriptstyle (2)}
-{\pi}_{\scriptscriptstyle (1)}^e\otimes{\pi}^d_{\scriptscriptstyle (2)})\biggr]\nonumber\\
&&+\omega^{dc}\partial_c\partial_aHF_{de}\otimes\frac{1}{2}\pi^a_{\scriptscriptstyle (1)}\otimes
\pi^e_{\scriptscriptstyle (2)}-\omega^{ec}\partial_c\partial_aH
F_{de}\otimes\frac{1}{2}\pi^a_{\scriptscriptstyle (1)}\otimes\pi^d_{\scriptscriptstyle (2)}\nonumber\\
&&+\omega^{ec}\partial_c\partial_aHF_{de}\otimes
\frac{1}{2}\pi^d_{\scriptscriptstyle (1)}\otimes\pi^a_{\scriptscriptstyle (2)}-\omega^{dc}\partial_c
\partial_aHF_{de}\otimes\frac{1}{2}\pi^e_{\scriptscriptstyle (1)}
\otimes\pi^a_{\scriptscriptstyle (2)}=\nonumber\\
&=&\omega^{ab}[\partial_bH\partial_aF_{de}+\partial_b\partial_dHF_{ae}+\partial_b\partial_eHF_{da}]\otimes
\frac{1}{2}(\pi^d_{\scriptscriptstyle (1)}\otimes\pi^e_{\scriptscriptstyle (2)}-
\pi^e_{\scriptscriptstyle (1)}\otimes\pi^d_{\scriptscriptstyle (2)}).
\end{eqnarray}

\bigskip

\noindent $\bullet$ \underline{\it Derivation of (\ref{comm1})}.

\begin{eqnarray}
[i\widehat{\cal H},\widehat{C}]&=&
[i\lambda_a\omega^{ab}\partial_bH\otimes {\bf 1}_{\scriptscriptstyle (1)}\otimes
{\bf 1}_{\scriptscriptstyle (2)}-i\omega^{be}\partial_e\partial_aH\otimes\pi^a_{\scriptscriptstyle (1)}
\xi_b^{\scriptscriptstyle (1)}\otimes {\bf 1}_{\scriptscriptstyle (2)}
\nonumber\\
&&-i\omega^{be}\partial_e\partial_aH\otimes
{\bf 1}_{\scriptscriptstyle (1)}\otimes\pi^a_{\scriptscriptstyle (2)}\xi_b^{\scriptscriptstyle (2)},
C_d\otimes\pi^d_{\scriptscriptstyle (1)}\otimes{\bf 1}_{\scriptscriptstyle (2)}
+C_d\otimes{\bf 1}_{\scriptscriptstyle (1)}\otimes\pi^d_{\scriptscriptstyle (2)}]=
\nonumber\\
&=&[i\lambda_a\omega^{ab}\partial_bH,C_d]\otimes \pi^d_{\scriptscriptstyle (1)}
\otimes{\bf 1}_{\scriptscriptstyle (2)}+[i\lambda_a\omega^{ab}\partial_bH,C_d]\otimes{\bf 1}
_{\scriptscriptstyle (1)}\otimes\pi^d
_{\scriptscriptstyle (2)}\nonumber\\
&&-i\omega^{be}\partial_e\partial_aHC_d\otimes[\pi^a_{\scriptscriptstyle (1)}\xi_b^{\scriptscriptstyle (1)},
\pi^d_{\scriptscriptstyle (1)}]\otimes{\bf 1}_{\scriptscriptstyle (2)}
-i\omega^{be}\partial_e\partial_aHC_d\otimes{\bf 1}_{\scriptscriptstyle (1)}
\otimes [\pi^a_{\scriptscriptstyle (2)}\xi_b^{\scriptscriptstyle (2)},\pi^d_{\scriptscriptstyle (2)}]=
\nonumber\\
&=&\partial_aC_d\omega^{ab}\partial_bH\otimes\pi^d_{\scriptscriptstyle (1)}\otimes
{\bf 1}_{\scriptscriptstyle (2)}+\partial_aC_d\omega^{ab}\partial_bH\otimes{\bf 1}_{\scriptscriptstyle (1)}
\otimes\pi^d_{\scriptscriptstyle (2)}\nonumber\\
&&+\omega^{ae}\partial_e\partial_dHC_a\otimes\pi^d_{\scriptscriptstyle (1)}\otimes{\bf 1}_{\scriptscriptstyle (2)}
+\omega^{ae}\partial_e\partial_dHC_a\otimes{\bf 1}_{\scriptscriptstyle (1)}\otimes
\pi^d_{\scriptscriptstyle (2)}=\nonumber\\
&=&(\partial_aC_d\omega^{ab}\partial_bH+\omega^{ae}\partial_e\partial_dHC_a)
\otimes[\pi^d_{\scriptscriptstyle (1)}\otimes{\bf 1}_{\scriptscriptstyle (2)}
+{\bf 1}_{\scriptscriptstyle (1)}\otimes\pi^d_{\scriptscriptstyle (2)}].
\end{eqnarray}

\bigskip

\noindent $\bullet$ \underline{\it Derivation of (\ref{genlie})}.

What we want to do now is to  convince the reader that, via the representation (\ref{5-9}) for the Lie derivative
$\widehat{\cal H}$ and with the representation (\ref{genform}) for a generic $m$-form
$\widehat{P}$, the commutator of $i\widehat{\cal H}$ with
$\widehat{P}$ gives just the action of the Lie derivative on 
$\widehat{P}$. First of all let us consider the following object 
\begin{equation}
\displaystyle \widehat{P}=\frac{1}{m!}P_{a_1\ldots a_m}(\varphi)
\otimes {\bf
A}\{\pi_{\scriptscriptstyle (1)}^{a_1}\otimes\pi_{\scriptscriptstyle (2)}^{a_2}\otimes\ldots
\pi_{\scriptscriptstyle (m)}^{a_m}\}\otimes{\bf 1}^{\otimes (2n-m)}, 
\end{equation}
where we have put all the operators $\pi$ in the first $m$ positions and antisymmetrized them by means of ${\bf A}$.
Let us calculate the commutator of $i\widehat{\cal H}$ with 
$\widehat{P}$:
\begin{eqnarray}
\displaystyle [i\widehat{\cal H}, \widehat{P}]&=&
\frac{1}{m!}[i\lambda_a\omega^{ab}\partial_bH,P_{a_1\ldots a_m}]\otimes{\bf A}
\{\pi_{\scriptscriptstyle (1)}^{a_1}\otimes\pi_{\scriptscriptstyle (2)}^{a_2}
\otimes\ldots\otimes \pi_{\scriptscriptstyle (m)}^{a_m}\}\otimes{\bf 1}^{\otimes (2n-m)}\nonumber\\
&&-\frac{i}{m!}P_{a_1\ldots a_m}\omega^{be}\partial_e\partial_aH\otimes {\bf A}\{[\pi_{\scriptscriptstyle (1)}^{a}
\xi^{\scriptscriptstyle (1)}_b,\pi_{\scriptscriptstyle (1)}^{a_1}]\otimes\ldots\otimes
\pi^{a_m}_{\scriptscriptstyle (m)}\}\otimes{\bf 1}^{\otimes (2n-m)}\nonumber\\
&&-\frac{i}{m!}P_{a_1\ldots a_m}\omega^{be}\partial_e\partial_aH\otimes {\bf A}\{\pi_{\scriptscriptstyle (1)}^{a_1}
\otimes[\pi_{\scriptscriptstyle (2)}^{a}
\xi^{\scriptscriptstyle (2)}_b,\pi_{\scriptscriptstyle (2)}^{a_2}]\ldots\otimes
\pi^{a_m}_{\scriptscriptstyle (m)}\}\otimes{\bf 1}^{\otimes (2n-m)}-\ldots\nonumber\\
&&-\frac{i}{m!}P_{a_1\ldots a_m}\omega^{be}\partial_e\partial_aH\otimes {\bf A}\{\pi_{\scriptscriptstyle (1)}^{a_1}
\ldots\otimes\pi_{\scriptscriptstyle (m-1)}^{a_{m-1}}\otimes[\pi_{\scriptscriptstyle (m)}^{a}
\xi^{\scriptscriptstyle (m)}_b,\pi_{\scriptscriptstyle (m)}^{a_m}]\}\otimes{\bf 1}^{\otimes (2n-m)}= \nonumber\\
&=&\frac{1}{m!}\omega^{ab}\partial_bH\partial_aP_{a_1\ldots a_m}\otimes{\bf A}
\{\pi_{\scriptscriptstyle (1)}^{a_1}\otimes\pi_{\scriptscriptstyle (2)}^{a_2}
\otimes\ldots\otimes \pi_{\scriptscriptstyle (m)}^{a_m}\}\otimes{\bf 1}^{\otimes (2n-m)}+\nonumber\\
&&+\frac{1}{m!}\omega^{a_1e}P_{a_1\ldots a_m}\partial_e\partial_aH\otimes{\bf A}
\{\pi_{\scriptscriptstyle (1)}^{a}\otimes\pi_{\scriptscriptstyle (2)}^{a_2}
\otimes\ldots\otimes \pi_{\scriptscriptstyle (m)}^{a_m}\}\otimes{\bf 1}^{\otimes (2n-m)}+\nonumber\\
&&+\frac{1}{m!}\omega^{a_2e}P_{a_1\ldots a_m}\partial_e\partial_aH\otimes{\bf A}
\{\pi_{\scriptscriptstyle (1)}^{a_1}\otimes\pi_{\scriptscriptstyle (2)}^{a}
\otimes\ldots\otimes \pi_{\scriptscriptstyle (m)}^{a_m}\}\otimes{\bf 1}^{\otimes (2n-m)}+\ldots\nonumber\\
&&+\frac{1}{m!}\omega^{a_me}P_{a_1\ldots a_m}\partial_e\partial_aH\otimes{\bf A}
\{\pi_{\scriptscriptstyle (1)}^{a_1}\otimes\pi_{\scriptscriptstyle (2)}^{a_2}
\otimes\ldots\otimes \pi_{\scriptscriptstyle (m)}^{a}\}\otimes{\bf 1}^{\otimes (2n-m)}.
\end{eqnarray}
If we now relabel the indices, we easily obtain:
\begin{eqnarray}
\displaystyle 
\displaystyle [i\widehat{\cal H}, \widehat{P}]&=&
\frac{1}{m!}\biggl[\omega^{ab}\partial_bH\partial_aP_{a_1\ldots a_m}+\sum_{i=1}^mP_{a_1\ldots\lambda\ldots a_m}
\omega^{\lambda e}\partial_e\partial_{a_i}H\biggr]\nonumber\\
&&\otimes {\bf A}\{\pi_{\scriptscriptstyle (1)}^{a_1}\otimes\pi_{\scriptscriptstyle (2)}^{a_2}
\otimes\ldots\otimes \pi_{\scriptscriptstyle (m)}^{a_m}\}\otimes {\bf 1}^{\otimes (2n-m)} \label{d4}
\end{eqnarray}
and this is just the way how an $m$-form transforms under the action of the Lie derivative \cite{Nakahara}:
\begin{equation}
{\cal L}_{(dH)^{\sharp}}P_{a_1a_2\ldots a_m}=
\omega^{ab}\partial_bH\partial_aP_{a_1\ldots a_m}+\sum_{i=1}^mP_{a_1\ldots\lambda\ldots a_m}
\omega^{\lambda e}\partial_e\partial_{a_i}H. \label{d5}
\end{equation}
The result (\ref{d4}) cannot depend on the position in which we place the operators $\pi$ inside the string of
the $2n$ Hilbert spaces. Therefore, even if we symmetrize in all the possible ways 
the operators $\pi$ and the identity operators ${\bf 1}$ constructing in this way the $m$-form given
by Eq. (\ref{genform}), we would obtain that the commutator of $i\widehat{\cal H}$ with
$\widehat{P}$ reproduces the correct action of the Lie derivative on the 
$m$-form given by Eq. (\ref{d5}).

\newpage

\section{}

\noindent In this Appendix we want to prove explicitly that the Hamiltonian $\widehat{\cal H}$ 
of Eq. (\ref{5-9}) is Hermitian under the scalar product (\ref{5-11}). This scalar product 
\begin{equation}
\displaystyle 
\langle \psi_1|\psi_2\rangle\equiv \int d^{2n}\varphi^a\prod_{i=1}^{2n}d^{2n}\pi_{\scriptscriptstyle
(i)}^a\,\psi_1^*\psi_2
\end{equation}
is formally identical to the usual scalar product of quantum mechanics. 
Therefore, like $q^i$ and $p_i$ in the usual quantum mechanics, also the operators $\varphi^a$,
$\lambda_a$, $\pi^a_{\scriptscriptstyle (i)}$ and $\xi_a^{\scriptscriptstyle (i)}$ are Hermitian under the scalar
product (\ref{5-11}). What we want to prove now is the hermiticity of the Hamiltonian (\ref{5-9}) and it goes 
as follows:
\begin{eqnarray}
\widehat{\cal H}^{\dagger}&=&(\lambda_a\omega^{ab}\partial_bH)^{\dagger}\otimes {\bf 1}^{\otimes 2n}-
\omega^{be}(\partial_e\partial_aH)^{\dagger}\otimes {\bf S}[(\pi^a\xi_b)^{\dagger}\otimes {\bf 1}^{\otimes (2n-1)}]=\nonumber\\
&=&\partial_bH\omega^{ab}\lambda_a\otimes{\bf 1}^{\otimes 2n}-
\omega^{be}\partial_e\partial_aH\otimes {\bf S}[\xi_b\pi^a\otimes
{\bf 1}^{\otimes (2n-1)}]=\nonumber\\
&=&(\lambda_a\omega^{ab}\partial_bH+i\partial_a\partial_bH\omega^{ab})
\otimes {\bf 1}^{\otimes 2n}-\omega^{be}\partial_e\partial_aH\otimes
{\bf S}[(\pi^a\xi_b+i\delta_b^a)\otimes {\bf 1}^{\otimes (2n-1)}]=\nonumber\\
&=&\lambda_a\omega^{ab}\partial_bH\otimes {\bf 1}^{\otimes 2n}-
\omega^{be}\partial_e\partial_aH\otimes {\bf S}[\pi^a\xi_b\otimes {\bf 1}^{\otimes (2n-1)}]=\nonumber\\
&=&\widehat{\cal H}
\end{eqnarray}
where we used the commutation relations $[\lambda_a,\partial_bH]=-i\partial_a\partial_bH$, 
$[\xi_a,\pi^l]=i\delta_a^l$ and the fact that $\partial_a\partial_bH\omega^{ab}=0$. 

\newpage

\section{}

In this Appendix we want to show that the expression (\ref{6-2}) of the Lie derivative leads to the standard 
transformations (\ref{4-6})-(\ref{4-8}) for vectors and forms via a proper choice of the generators $G^a_{\;b}$.
Let us use the expression (\ref{6-4}) instead of (\ref{6-2}) and let us make it act on vectors with components 
$V^e$:
\begin{equation}
\displaystyle {\cal L}_hV^e=h^f\partial_fV^e+\frac{i}{2}K_{ab}(\Sigma^{ab}_{\text{vec}})^e_{\;f}V^f.
\label{e-1}
\end{equation}
We have put the indication ``$\text{vec}$" on $\Sigma^{ab}$ to indicate that we have to choose the vector representation
of the operator $\Sigma^{ab}$. The indices $e$ and $f$ are matrix indices while $(a,b)$ are group (or algebra) indices.
We know how a vector is transformed under a Lie derivative (see Eq. \ref{4-6})
\begin{equation}
{\cal L}_hV^e=h^f\partial_fV^e-\partial_fh^eV^f \label{e-2}
\end{equation}
and comparing (\ref{e-1}) and (\ref{e-2}) we get that 
\begin{equation}
(\Sigma^{ab}_{\text{vec}})^e_{f}=-i(\delta^a_{\; f}\omega^{be}+\delta^b_f\omega^{ae}). \label{e-3-a}
\end{equation}
Let us now see what we get by inserting this expression in the operator $S$ of (\ref{6-5}) which gives us the infinitesimal
$\text{Sp}(2N)$ transformation in the tangent space. The result is
\begin{equation}
\displaystyle S^e_{\;f}=\biggl(1-\frac{i}{2}K_{ab}\Sigma^{ab}_{\text{vec}}\biggr)^e_{\;f}=
\delta^e_f+\omega^{ea}K_{af} \label{e-3}
\end{equation}
and this is exactly \cite{littlejohn} the expression for the infinitesimal $\text{Sp}(2N)$ transformation for vectors.
The same kind of steps can be done for forms and in that case Eq. (\ref{e-1}) is replaced by 
\begin{equation}
\displaystyle {\cal L}_h\alpha_f=h^a\partial_a\alpha_f+\frac{i}{2}
K_{ab}(\Sigma^{ab}_{\text{form}})_f^{\;\;e}\alpha_e \label{e-4}
\end{equation}
where $\alpha_f$ are the coefficients of forms.
This equation must be equal to the standard action of the Lie derivative on forms which is
\begin{equation}
{\cal L}_h\alpha_f=h^a\partial_a\alpha_f+\partial_fh^e\alpha_e \label{e-5}
\end{equation}
and comparing (\ref{e-4}) with (\ref{e-5}) we get 
\begin{equation}
(\Sigma^{ab}_{\text{form}})^{\;\;e}_f=i(\delta^a_f\omega^{be}+\delta^b_f\omega^{ae}).
\end{equation}
From this $(\Sigma^{ab}_{\text{form}})$ we can easily get the matrix $S_f^{\;\;e}$ which is
\begin{equation}
S_f^{\;\;e}=\delta_f^{\;\;e}-\omega^{ea}K_{af}
\end{equation}
and this is exactly \cite{littlejohn} the expression for the infinitesimal $\text{Sp}(2N)$ transformation for forms.
The reader may be puzzled by the fact that in all this Appendix no Greek $(\alpha,\beta)$ indices have appeared
to indicate the representation but only Latin indices. The reason is that the vector and form representation have the same
dimension as the space or the algebra of the group and so the $\alpha$-indices are as many as the $a$-indices and we have
indicated them with the same notation. This will not be the case for other representations.

\newpage

\section{}

In this Appendix we shall show that the composite object defined in (\ref{7-9}) has the same equation of motion as the 
Jacobi fields (\ref{7-8}). Suppressing, when it is not necessary, the indices $x$ and $y$ and using the equations of motion
for $\eta^x$ and $\bar{\eta}_y$, let us do the time derivative of the LHS and the RHS of Eq. (\ref{7-9})
\begin{eqnarray}
\displaystyle \partial_t{\mathscr P}^d&=&(\partial_t\bar{\eta})\gamma^d\eta+\bar{\eta}\gamma^d(\partial_t\eta)=
\nonumber\\
&=&\frac{i}{2}\partial_a\partial_bH\bar{\eta}_y(\Sigma^{ab}_{\text{meta}})^y_{\;x}(\gamma^d)^x_{\;z}\eta^z-
\frac{i}{2}\bar{\eta}_x(\gamma^d)^x_{\;z}\partial_a\partial_bH(\Sigma_{\text{meta}}^{ab})^z_{\;y}\eta^y=\nonumber\\
&=&\frac{i}{2}\partial_a\partial_bH\biggl\{\bar{\eta}_x\bigl[(\Sigma_{\text{meta}}^{ab})^x_{\;z}(\gamma^d)^z_{\;y}-
(\gamma^d)^x_{\;z}(\Sigma_{\text{meta}}^{ab})^z_{\;y}\bigr]\eta^y\biggr\}. \label{f-1}
\end{eqnarray}
The expression inside the square bracket on the RHS is the commutator of $\Sigma_{\text{meta}}$ and $\gamma$ that can be worked out using
the expression (\ref{6-9-a}) of $\Sigma_{\text{meta}}$ and the commutators (\ref{6-7}) among the $\gamma^a$.
The result is
\begin{equation}
[\Sigma_{\text{meta}}^{ab},\gamma^d]^x_{\;y}=i[\omega^{ad}\gamma^b+\omega^{bd}\gamma^a]^x_{\;y}. \label{f-2}
\end{equation}
Inserting (\ref{f-2}) in (\ref{f-1}) we get 
\begin{equation}
\displaystyle \partial_t{\mathscr P}^d=-\frac{1}{2}\partial_a\partial_bH
\bigl[\bar{\eta}_x(\omega^{ad}\gamma^b+\omega^{bd}\gamma^a)^x_{\;y}\eta^y\bigr]. \label{f-3}
\end{equation}
On the RHS of this expression we can easily recognize combinations of $\bar{\eta},\gamma,\eta$ which reproduce expression
(\ref{7-9}) so (\ref{f-3}) can be rewritten as
\begin{eqnarray}
\displaystyle \partial_t{\mathscr P}^d&=&-\frac{1}{2}\partial_a\partial_bH[\omega^{ad}{\mathscr P}^b
+\omega^{bd}{\mathscr P}^a]=\nonumber\\
&=&\omega^{da}\partial_a\partial_bH{\mathscr P}^b\nonumber\\
&=&\partial_bh^d{\mathscr P}^b \label{f-4}
\end{eqnarray}
and this is exactly the equation of motion (\ref{7-8}) of the Jacobi fields, so ${\mathscr P}^a=\delta{\cal \varphi}^a$.

\newpage

\section{}

In this Appendix we shall show that in the expression (\ref{8-4}), if we transform the 
$|\psi^{\scriptscriptstyle (p)}\rangle$
according to the metaplectic representation, then the $\rho$ will turn out to get transformed according the symplectic one.
Let us limit ourselves to the two-form
\begin{equation}
\displaystyle
\rho_{ab}=\frac{1}{2}\text{Tr}\biggl[|\psi^{\scriptscriptstyle (2)}
\rangle\langle\psi^{\scriptscriptstyle (2)}|(\gamma_a\otimes\gamma_b-\gamma_b\otimes\gamma_a)\biggr]
\end{equation}
in components it means
\begin{equation}
\displaystyle \rho_{ab}=\frac{1}{2}\psi_{\scriptscriptstyle (2)}^{xy}
\psi_{zw}^{\scriptscriptstyle (2)}(\gamma_a)^z_{\;x}(\gamma_b)^w_{\;y}
-\frac{1}{2}(a\leftrightarrow b).
\end{equation}
Transforming the $\psi^{xy}$ according to the metaplectic representation (\ref{6-9}) and making use of (\ref{6-8}) we get
\begin{eqnarray}
\displaystyle \rho_{ab}^{\prime}&=&\frac{1}{2}\psi_{\scriptscriptstyle (2)}
^{\prime xy}\psi_{zw}^{\prime {\scriptscriptstyle (2)}}
(\gamma_a)^z_{\;x}(\gamma_b)^w_{\;y}-\frac{1}{2}(a\leftrightarrow b)=\nonumber\\
&=&\frac{1}{2}M^x_{\;t}M^y_{\;s}\psi_{\scriptscriptstyle (2)}^{ts}
\psi_{uv}^{\scriptscriptstyle (2)}M^{\dagger u}_{\;\;\;z}M^{\dagger v}_{\;\;\;w}
(\gamma_a)^z_{\;x}(\gamma_b)^w_{\;y}-\frac{1}{2}(a\leftrightarrow b)=\nonumber\\
&=&\frac{1}{2}\psi_{\scriptscriptstyle (2)}^{ts}\psi_{uv}^{\scriptscriptstyle (2)}M^{\dagger u}_{\;\;\;z}M^{\dagger v}_{\;\;\;w}(\gamma_a)^z_{\;x}
(\gamma_b)^w_{\;y}M^x_{\;t}M^y_{\;s}-\frac{1}{2}(a\leftrightarrow b)=\nonumber\\
&=&\frac{1}{2}\psi_{\scriptscriptstyle (2)}^{ts}\psi_{uv}^{\scriptscriptstyle (2)}
(\gamma_d)^u_{\;t}S^d_{\;a}(\gamma_f)^v_{\;s}S^f_{\;b}-
\frac{1}{2}(a\leftrightarrow b)=\nonumber\\
&=&\frac{1}{2}\rho_{df}S^d_{\;a}S^f_{\;b}-\frac{1}{2}\rho_{df}S^d_{\;b}S^f_{\;a}=\nonumber\\
&=&\rho_{df}S^d_{\;a}S^f_{\;b}.
\end{eqnarray}
This proves that $\rho_{ab}$ transforms according to the symplectic representation of forms.

\end{document}